\documentclass[a4paper,twocolumn,11pt, floatfix, accepted=2026-04-20]{quantumarticle}
\pdfoutput=1
\usepackage{amsmath}
\usepackage{amssymb}
\usepackage[utf8]{inputenc}
\usepackage[english]{babel}
\usepackage[T1]{fontenc}
\usepackage{hyperref}
\usepackage{ragged2e}
\usepackage{mathtools}
\usepackage{graphicx}% Include figure files
\usepackage[caption=false]{subfig}
\usepackage{dcolumn}% Align table columns on decimal point
\usepackage{xcolor}
\usepackage{bm}% bold math
\usepackage{bbold}
\usepackage{braket}
\usepackage{siunitx} % For proper display of units
\usepackage[capitalize]{cleveref}
\usepackage[numbers,sort&compress]{natbib}
\usepackage{hyperref}

\DeclareCaptionJustification{justified}{\justifying}

\newcommand{\ie}{i.\,e.\ }

\newcommand{\eg}{e.\,g.\ }

\DeclarePairedDelimiter\abs{\lvert}{\rvert}%
\DeclarePairedDelimiter\expect{\langle}{\rangle}%

\crefname{section}{Sec.}{Sec.}
\crefname{appendix}{App.}{App.}

\begin{document}

\title{Learning symmetry-protected topological order from trapped-ion experiments}

\author{Nicolas Sadoune}
\affiliation{\mbox{Arnold Sommerfeld Center for Theoretical Physics, Ludwig-Maximilians-Universit\"{a}t M\"{u}nchen, 80333 M\"{u}nchen, Germany}}
\affiliation{Munich Center for Quantum Science and Technology, 80799 M\"{u}nchen, Germany}

\author{Ivan Pogorelov}
\affiliation{Institute for Experimental Physics, University of Innsbruck, 6020 Innsbruck, Austria}

\author{Claire L. Edmunds}
\affiliation{Institute for Experimental Physics, University of Innsbruck, 6020 Innsbruck, Austria}

\author{Giuliano Giudici}
\affiliation{Institute for Theoretical Physics, University of Innsbruck, 6020 Innsbruck, Austria}
\affiliation{Institute for Quantum Optics and Quantum Information of the Austrian Academy of Sciences, 6020 Innsbruck, Austria}
\affiliation{PlanQC GmbH, 85748 Garching, Germany}
\affiliation{\mbox{Arnold Sommerfeld Center for Theoretical Physics, Ludwig-Maximilians-Universit\"{a}t M\"{u}nchen, 80333 M\"{u}nchen, Germany}}
\affiliation{Munich Center for Quantum Science and Technology, 80799 M\"{u}nchen, Germany}

\author{Giacomo Giudice}
\affiliation{PlanQC GmbH, 85748 Garching, Germany}

\author{Christian D. Marciniak}
\affiliation{Institute for Experimental Physics, University of Innsbruck, 6020 Innsbruck, Austria}

\author{Martin Ringbauer}
\affiliation{Institute for Experimental Physics, University of Innsbruck, 6020 Innsbruck, Austria}

\author{Thomas Monz}
\affiliation{Institute for Experimental Physics, University of Innsbruck, 6020 Innsbruck, Austria}
\affiliation{Alpine Quantum Technologies GmbH, 6020 Innsbruck, Austria}

\author{Lode Pollet}
\affiliation{\mbox{Arnold Sommerfeld Center for Theoretical Physics, Ludwig-Maximilians-Universit\"{a}t M\"{u}nchen, 80333 M\"{u}nchen, Germany}}
\affiliation{Munich Center for Quantum Science and Technology, 80799 M\"{u}nchen, Germany}

%\date{\today}

\begin{abstract}
Classical machine learning has proven remarkably useful in post-processing quantum data, yet typical learning algorithms often require prior training to be effective. In this work, we employ a tensorial kernel support vector machine (TK-SVM) to analyze experimental data produced by trapped-ion quantum computers. This unsupervised method benefits from directly interpretable training parameters, allowing it to identify the non-trivial string-order characterizing symmetry-protected topological (SPT) phases. We apply our technique to two examples: a spin-1/2 model and a spin-1 model, featuring the cluster state and the AKLT state as paradigmatic instances of SPT order, respectively. Using matrix product states, we generate a family of quantum circuits that host a trivial phase and an SPT phase, with a sharp phase transition between them. For the spin-1 case, we implement these circuits on two distinct trapped-ion machines based on qubits and qutrits. Our results demonstrate that the TK-SVM method successfully distinguishes the two phases across all noisy experimental datasets, highlighting its robustness and effectiveness in quantum data interpretation.
\end{abstract}

\maketitle

\section{Introduction}
Present-day noisy intermediate-scale quantum (NISQ) devices~\cite{NISQ_Preskill2018} have shown remarkable progress in the coherent control of several qubits via high-fidelity quantum operations and efficient readout. 
Various hardware platforms are being concurrently developed, based \eg on superconducting qubits~\cite{annurev_qubit2019, CQED_RMP_2021,Satzinger2021,IBM2023}, neutral atoms~\cite{Saffman_2016,Levine2019,Pasqal2020,Bluvstein2022}, and trapped ions~\cite{CiracZoller2000,postler2022,moses2023race,kang2023quantum}.
Despite significant advances in active error correction~\cite{postler2022,Bravyi2024,Bluvstein2024,dasilva2024}, all these platforms remain error-prone and limited in size, with only a few high-quality qubits at disposal. 
Therefore, it is of fundamental interest to understand whether NISQ devices can be of practical use in a pre-fault-tolerant era.

Here, we tackle this question by applying efficient classical machine learning methods~\cite{HuangPreskill_provably_2022} to the many-body problem of phase classification of quantum data.
Specifically, we analyze two distinct datasets obtained from two trapped-ion machines: one based on qubits~\cite{pogorelov2021} and the other on qutrits~\cite{Ringbauer2022}. Such platforms feature high-fidelity gate operations and universal connectivity~\cite{Srinivas2021,Clark2021}, enabling the realization of a wide variety of quantum circuits.
We follow the proposal of Ref.~\cite{sadoune23}, in which synthetic data of a cluster Ising model were analyzed~\cite{Raussendof2001,Nielsen2004,Verstraete2004,Nielsen2006,Smacchia2011}, and extend it to the technically more challenging case of a spin-1 model that includes the Affleck-Kennedy-Lieb-Tasaki (AKLT) state~\cite{AKLT}.
Our scheme combines ideas from positive-operator valued measurements (POVM)~\cite{BookNielsenChuang}, tensor network states~\cite{Schollwoeck2011}, shadow tomography~\cite{Aaronson2018,Aaronson2019}, and unsupervised interpretable machine learning~\cite{Greitemann_2019,Liu_2019}. 
Despite this method being expected to perform better on large system sizes, we demonstrate its capabilities on noisy experimental datasets extracted from 8 qubits and 5 qutrits, by successfully identifying symmetry-protected topological order in spin-1/2 and spin-1 systems.

\section{Motivation and Strategy}\label{sec:strategy}
Quantum advantage involves utilizing NISQ devices as quantum simulators for systems too large to be simulated classically, thus shifting the theoretical focus from providing predictions via state-of-the-art many-body numerical methods to extracting valuable information from datasets obtained through a limited number of quantum measurements.
Classical machine learning algorithms are well-suited for this task. However, despite significant recent progress in the field, these approaches often remain heuristic and hard to interpret.
An important step forward was achieved in Ref.~\cite{HuangPreskill_provably_2022}: For a certain class of problems, it was proven that classical machine learning algorithms that learn from quantum data are efficient, meaning the required training data scales polynomially with the size of the quantum system under investigation. 
For example, the problem of classifying many-body phases can be solved efficiently, provided that the corresponding order parameters are sufficiently local~\footnote{Thus excluding e.g. topologically ordered phases such as the one exhibiting the fractional quantum Hall effect.}.

In fact, the size of typical quantum datasets is too small for training sophisticated deep learning models for \eg full tomography of a strongly correlated many-body state~\cite{Torlai2018,Carrasquilla2019}.
Yet, quantum phase classification is a simpler task that can be accomplished with less expressive classical machine learning algorithms.
The main purpose of the present work is to show that such a task is achievable in the NISQ era on state-of-the-art trapped-ion machines, as well as augment the results of Ref.~\cite{HuangPreskill_provably_2022} by eliminating the requirement of prior labeling of the training data~\cite{sadoune2024}. 

A machine learning algorithm well apt to this task is the tensorial-kernel support vector machine (TK-SVM)~\cite{Ponte17,Greitemann_2019,Liu_2019, Greitemann_PhD:2019}. In our work, this consists of a linear-kernel SVM~\cite{BookVapnik} which takes as input tensors constructed from real-space snapshots of a quantum many-body system. 
From the resulting tensorial kernel, TK-SVM inherits its {\it interpretable} character: The structure constants of the decision function encode the order parameter of the quantum phase~\footnote{A local constraint as found in lattice gauge theories or in classical spin liquids can also be learned, cf. Ref~\cite{Greitemann19b}}. Moreover, by analyzing the bias term of the decision function, in combination with graph-theoretical considerations, it is possible to make this method quasi-unsupervised~\cite{Greitemann_2019, Liu_2019, Greitemann_PhD:2019}, eliminating the need for prior training and enabling the detection of phases that are unknown a priori. Indeed, TK-SVM has identified novel phases with large unit cells in classical Kitaev magnets~\cite{Liu_2021} or with emergent subsystem symmetries on the breathing pyrochlore lattice~\cite{sadoune2024}. The extension of TK-SVM to quantum systems requires dealing with non-commuting observables, which can be accomplished through informationally complete POVM measurements~\cite{sadoune23}.

A challenging question in quantum phase classification is how to discriminate between symmetry-protected topological (SPT) phases with string orders and topologically trivial phases.
Addressing this problem on a trapped-ion machine requires preparing many-body states that lie in such phases and performing measurements within the limits of current state-of-the-art technology.
In our experiments, the primary limiting factor is the maximum circuit depth achievable without significant loss of fidelity (see Appendix~\ref{app:CircuitDetails}).
For this reason, we focus on matrix product state (MPS) models with bond dimension 2, which can be prepared using shallow circuits with a depth that scales linearly with the system size. These circuits can be reliably implemented on our qubit and qutrit machines for chains of up to 8 qubits and 5 qutrits, respectively.

The rest of the paper is structured as follows.
In Sec.~\ref{sec:models} we introduce the MPS representations of the families of spin-$1/2$ and spin-$1$ states that we studied.
In Sec.~\ref{sec:StatePrep} we show how these are implemented on the trapped-ion qubit and qutrit hardware, followed by the measurement protocol in Sec.~\ref{sec:measurement}. The experimental setup is discussed in Sec.~\ref{sec:experimental_setup}. The machine learning algorithm employed for extracting information from snapshots of spin-$1/2$ and spin-$1$ models is outlined in detail in Sec.~\ref{sec:ML}. The results are presented in Sec.~\ref{sec:results}, followed by an outlook and conclusion in Sec.~\ref{sec:outlook}. In the appendices we give detailed information on circuit transpilation, a comparison between the qubit and qutrit implementation, and an analysis of the scaling properties of our scheme.

\section{Investigated quantum states}\label{sec:models}
We consider two related families of quantum states. Each family corresponds to a parametrized set of states that have a MPS representation with low bond dimension. One family comprises a subset of ground states of a spin-$1/2$ cluster Ising model~\cite{Smith22}, whereas the other family consists of spin-$1$ states that contains the AKLT state.

\subsection{Family of Spin-1/2 States}
We consider translation invariant matrix product states of the form
\begin{equation}\label{eq:cluster-mps}
\ket{\psi(g)} \propto \sum_{ s_1 \dots s_L } \mathrm{Tr} \left[ B_{s_1}(g) \dots B_{s_L}(g) \right] \ket{s_1 \dots s_L},
\end{equation}
where $L$ is the number of lattice sites, and $g$ is a real parameter to be defined below.
For a spin-1/2 system $s=0,1$ labels the spin-$1/2$ or qubit computational basis states. 
The MPS tensors for the spin-1/2 model that we analyze are~\cite{Wolf06}
\begin{equation}
\label{eq:cluster-tensor}
B_0 = \begin{pmatrix}
0 & 0 \\
1 & 1
\end{pmatrix},\quad
B_1 = \begin{pmatrix}
1 & g \\
0 & 0
\end{pmatrix} \, .
\end{equation}
This MPS interpolates between the cluster state ($g=1$) and a trivial product state ($g=-1$), and it satisfies the $\mathbb{Z}_2 \times \mathbb{Z}_2$ symmetry generated by $\prod_{i=1}^{L/2} X_{2 i}$ and $\prod_{i=1}^{L/2} X_{2 i-1}$, where $X,Y,Z$ are the standard Pauli matrices.
A phase transition occurs at $g=0$, separating an SPT phase from a trivial phase~\cite{Wolf06}.
The SPT phase can be characterized by the following string order parameter~\cite{perez-garcia2008}
\begin{equation}
\begin{split}
\lim_{\ell \to \infty} \! \Braket{\! Z_i Y_{i+1}
\left(\prod_{j=i+2}^{\ell - 2} X_j \right)
Y_{\ell - 1} Z_\ell \!}
= {} & \\
\begin{cases}
    \frac{-4 g}{(1-g)^2} \quad & \mbox{if}~g < 0\\
    0 \quad & \mbox{if}~g \geq 0,
\end{cases}
\end{split}
\end{equation}
%\begin{equation}
%    \lim_{\ell \to \infty} \! \Braket{\! Z_i Y_{i+1} \left(\prod_{j=i+2}^{\ell - 2} X_j \right) Y_{\ell - 1} Z_\ell \!} = 
%    \begin{cases}
%        \frac{-4 g}{(1-g)^2}\quad &\mbox{if}~g < 0\\
%        0 \quad &\mbox{if}~g \geq 0,
%    \end{cases}
%\end{equation}
where the index $i$ labels any site of an infinite chain.

\begin{figure}
\centering
\includegraphics[width=0.9\columnwidth]{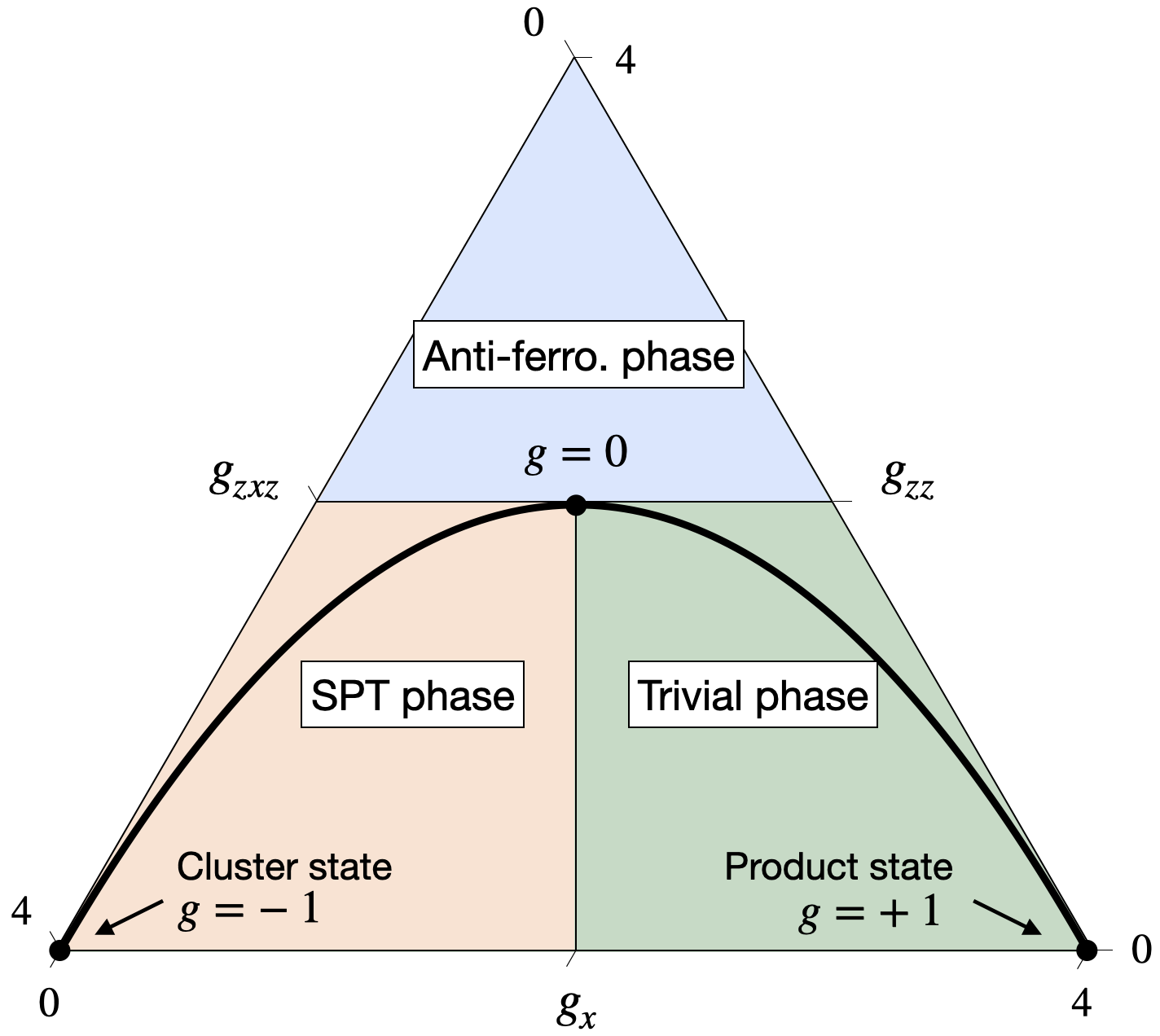}
\caption{Phase diagram of the 3-parameter cluster Ising model Eq.~\eqref{eq:cluster-hamiltonian}, featuring a MPS path from the SPT phase to the trivial paramagnetic phase. On the black line $g_{zxz} = (g-1)^2$, $g_{zz} = 2 ( 1- g^2)$, $g_x = (1+g)^2$ the ground state is the MPS defined by the tensors in Eq.~\eqref{eq:cluster-tensor}.}
\label{fig:mpsLine_clusterModel}
\end{figure}

The MPS in Eq.~\eqref{eq:cluster-mps} is the ground state of the Hamiltonian
\begin{equation}
\label{eq:cluster-hamiltonian}
H=\sum_j g_{zxz} Z_j X_{j+1} Z_{j+2} -g_{zz} Z_j Z_{j+1} - g_x X_j \, ,
\end{equation}
where the sum runs over the lattice sites and the relation between the MPS parameter $g \in [-1,1]$ and the Hamiltonian coupling constants is $g_{zxz} = (g-1)^2, g_{zz} = 2(1-g^2), g_x=(1+g)^2$.
This Hamiltonian exhibits three phases: A topological phase protected by a $\mathbb{Z}_2\times \mathbb{Z}_2$ symmetry (SPT), a trivial paramagnetic phase, and a symmetry-broken anti-ferromagnetic phase. The MPS line goes through the former two phases, and its phase transition point corresponds to the tricritical point where all three phases meet (see Fig.~\ref{fig:mpsLine_clusterModel}).
We stress that the simplicity of the MPS parent Hamiltonian is not instrumental to the analysis that follows. In fact, we will always consider states along the MPS line since they can be prepared via shallow circuits, as we illustrate in Sec.~\ref{sec:StatePrep}. 

\subsection{Family of Spin-1 States}
We aim to find a family of states with the same properties as the spin-$1/2$ MPS model in Eqs.~\eqref{eq:cluster-mps}-\eqref{eq:cluster-tensor}, but living on a spin-$1$ Hilbert space. To this end, we consider the isometry $P$ which maps two consecutive spin-$1/2$ to a single spin-$1$ degree of freedom
\begin{equation}
    \begin{split}
        \frac{\ket{X_+ X_-} \pm i \ket{X_- X_+}}{\sqrt{2}} &\to \ket{\pm} , \\
        \ket{X_+ X_+} &\to \ket{\circ} ,
    \end{split}
\label{eq:map_P}
\end{equation}
where $+,\circ,-$ label the qutrit computational basis states, and $\ket{X_\pm}$ are the eigenstates of $X$. When applied to the cluster state (Eqs.~\eqref{eq:cluster-mps}-\eqref{eq:cluster-tensor} for $g=-1$) this transformation yields the well-known AKLT state~\cite{Verresen17,AKLT}, which is the prototypical example of the same SPT order encoded in the cluster state but realized on a qutrit system. Thanks to the invariance under $SO(3)$, its parent Hamiltonian takes the simple form
\begin{equation}
H = \sum_j \bm{S}_j \cdot\bm{S}_{j+1} + \frac{1}{3} ( \bm{S}_j \cdot \bm{S}_{j+1})^2.
\label{eq:aklt_ham}
\end{equation}
where $\bm{S} = (\tau_x,\tau_y,\tau_z)$ are the spin-1 matrices 
\begin{equation}\label{eq:qutrit-paulis}
\resizebox{0.48\textwidth}{!}{$
\tau^x = \frac{1}{\sqrt{2}}\begin{pmatrix}
0 & 1 & 0 \\
1 & 0 & 1 \\
0 & 1 & 0
\end{pmatrix}\!,
\tau^y = \frac{1}{\sqrt{2}}\begin{pmatrix}
0 & -i & 0 \\
i & 0 & -i \\
0 & i & 0
\end{pmatrix}\!,
\tau^z = \begin{pmatrix}
1 & 0 & 0 \\
0 & 0 & 0 \\
0 & 0 & -1
\end{pmatrix}
$}.
\end{equation}

We can extend this mapping to the whole MPS line parametrized by Eqs.~\eqref{eq:cluster-mps}-\eqref{eq:cluster-tensor} and obtain a family of spin-$1$ states with the following MPS tensors:
\begin{equation}
\label{eq:qutrit-mps}
\begin{split}
B_+ &= \frac{1}{\sqrt{2}}
\begin{pmatrix}
\frac{1-g}{2} + i\frac{1+g}{2} & ig \\
-i & \frac{g-1}{2} - i\frac{1+g}{2}
\end{pmatrix},\\
B_\circ &= \begin{pmatrix}
\frac{1+g}{2} & g\\
1 & \frac{1+g}{2}
\end{pmatrix},\\
B_- &= B_+^\ast.
\end{split}
\end{equation}
For $g=-1$, the standard MPS representation of the AKLT state~\cite{schollwock2011} is recovered via the gauge transformation $\frac{1}{\sqrt{2}}\begin{psmallmatrix} i & -1 \\ i & 1 \end{psmallmatrix}$. For $g = 1$, this MPS is the product state $\bigotimes \ket{\circ}$. For all $g$ its transfer matrix can be diagonalized analytically, and all local quantities can be expressed in closed form. For instance, the string order parameter characterizing the SPT phase reads
\begin{equation}
    \lim_{\ell \to \infty} \Braket{\tau^z_i \prod_{j=i+1}^{\ell-1} e^{i \pi \tau^z_j} \tau_\ell^z} = 
    \begin{cases}
        -\frac{16}{9}\frac{g}{(1-g)^2}\quad &\mbox{if}~g < 0\\
        0 \quad &\mbox{if}~g \geq 0.
    \end{cases}
\end{equation}
As was the case for the spin-$1/2$ MPS in Eqs.~\eqref{eq:cluster-mps}-\eqref{eq:cluster-tensor}, the string order parameter is nonzero for $g < 0$ and vanishes at $g=0$, where a phase transition to a trivial phase occurs. 
For $g \ne -1$ the MPS realized by the tensors in Eq.~\eqref{eq:qutrit-mps} is not $SO(3)$-invariant as the AKLT state. For this reason, its parent Hamiltonian is more complicated than Eq.~\eqref{eq:aklt_ham} and we omit its expression.

\section{State preparation}\label{sec:StatePrep}
\begin{figure}
\centering
\includegraphics[width=0.95\columnwidth]{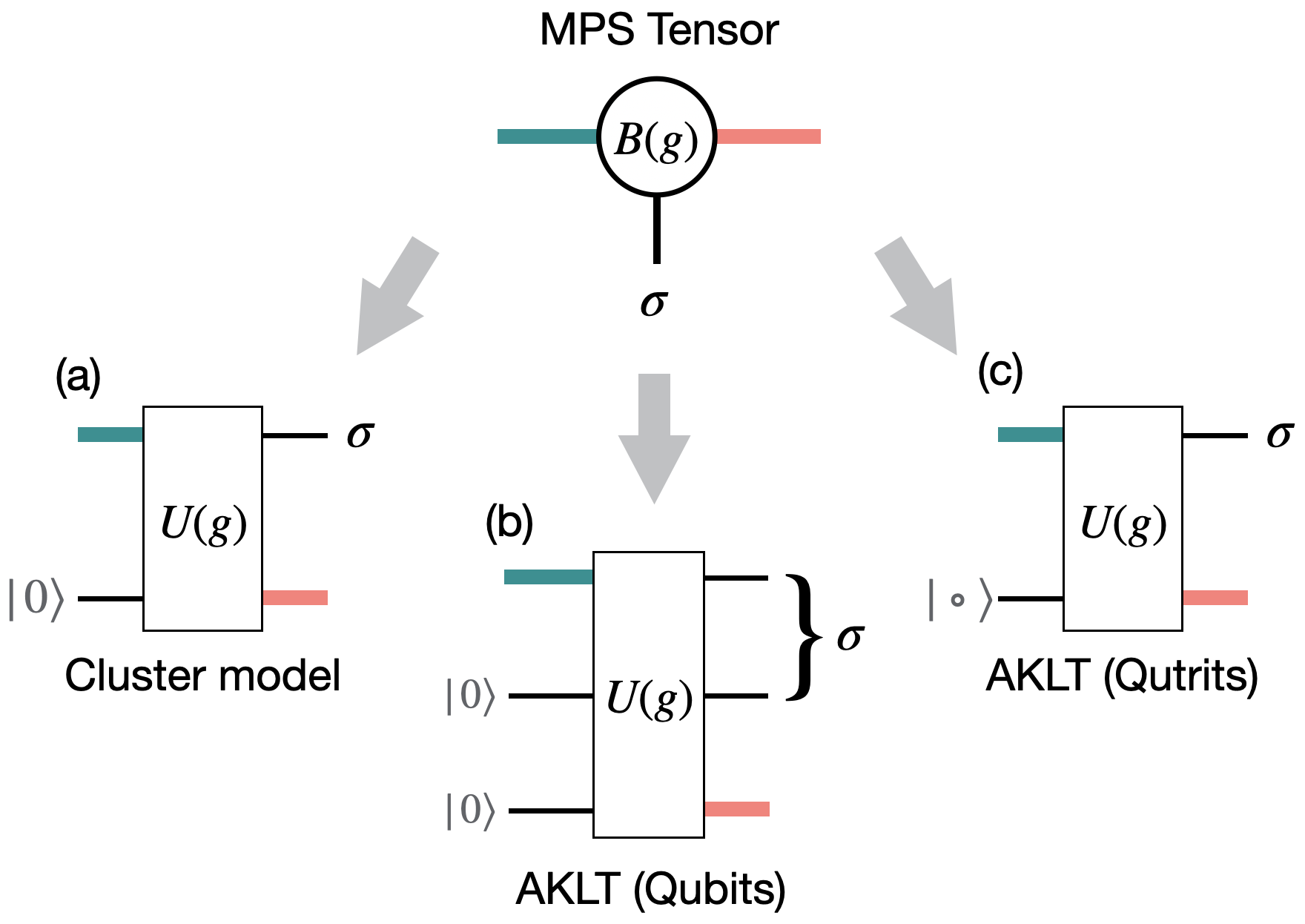}
\caption{\textbf{From MPS tensors to unitary gates.} The MPS tensor is converted into a gate by first reshaping the tensor, which is seen as an isometry from the left virtual space (green leg) to the product of right virtual and physical spaces (orange and black legs). A dummy leg of the proper dimension is then added to the tensor to make it a square matrix, which can be unitarized with standard routines \cite{Iten21,Younis21}.  
For the spin-$1/2$ case the resulting gate acts on 2 qubits (a). For the spin-$1$ case, instead, two different implementations are possibile, that turn the MPS into a 3-qubit gate (b) or into a 2-qutrit gate (c). See the text for more details.
}
\label{fig:Unitarization}
\end{figure}

The low bond dimension of the MPS families discussed above enables a simple and resource-efficient state preparation in gate-based quantum computing platforms. We follow the procedure given in Ref.~\cite{Barratt21} to prepare an infinite-size translational invariant MPS via a finite-depth quantum circuit with local unitaries $U(g)$. This procedure is schematically depicted in Fig.~\ref{fig:Unitarization}. The MPS tensor left and right virtual indices are the green and orange legs, while the physical index is the black leg.
The first step consists of transforming the MPS into right-canonical form and interpreting the right-normalized tensor as an isometry from the left virtual space (green leg in Fig.~\ref{fig:Unitarization}) to the product of physical and right virtual spaces (black and orange legs in Fig.~\ref{fig:Unitarization}). 
The isometry is then transformed into a unitary matrix by adding an extra leg with the same dimension of the physical space. 
This leg is equivalent to a dummy index for the MPS tensor when applied to the state $\ket{0}$ for the cluster state case and $\ket{0 0}$ (or $\ket{\circ}$) for the AKLT case (cf. Fig.~\ref{fig:Unitarization}), but its action on the remaining physical states is free and adjustable to make the gate $U(g)$ unitary. 

\begin{figure}
\centering
\subfloat[Spin-$1/2$ family \label{subfig:CircuitCluster}]
{\includegraphics[width=0.9\columnwidth]{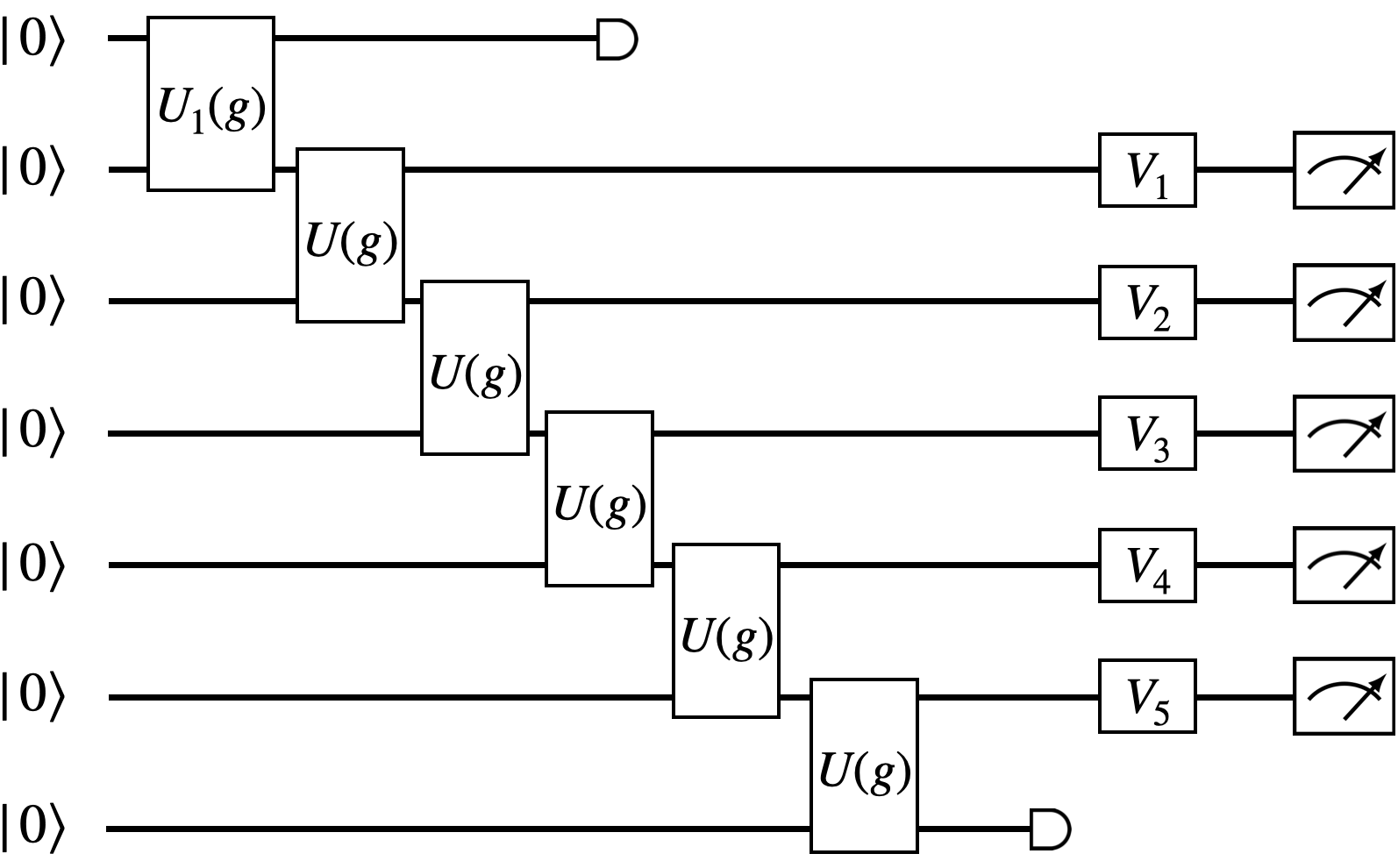}}\\[0.5cm]
\subfloat[Spin-$1$ family  (qubits)\label{subfig:CircuitAKLTqubits}]
{\includegraphics[width=0.9\columnwidth]{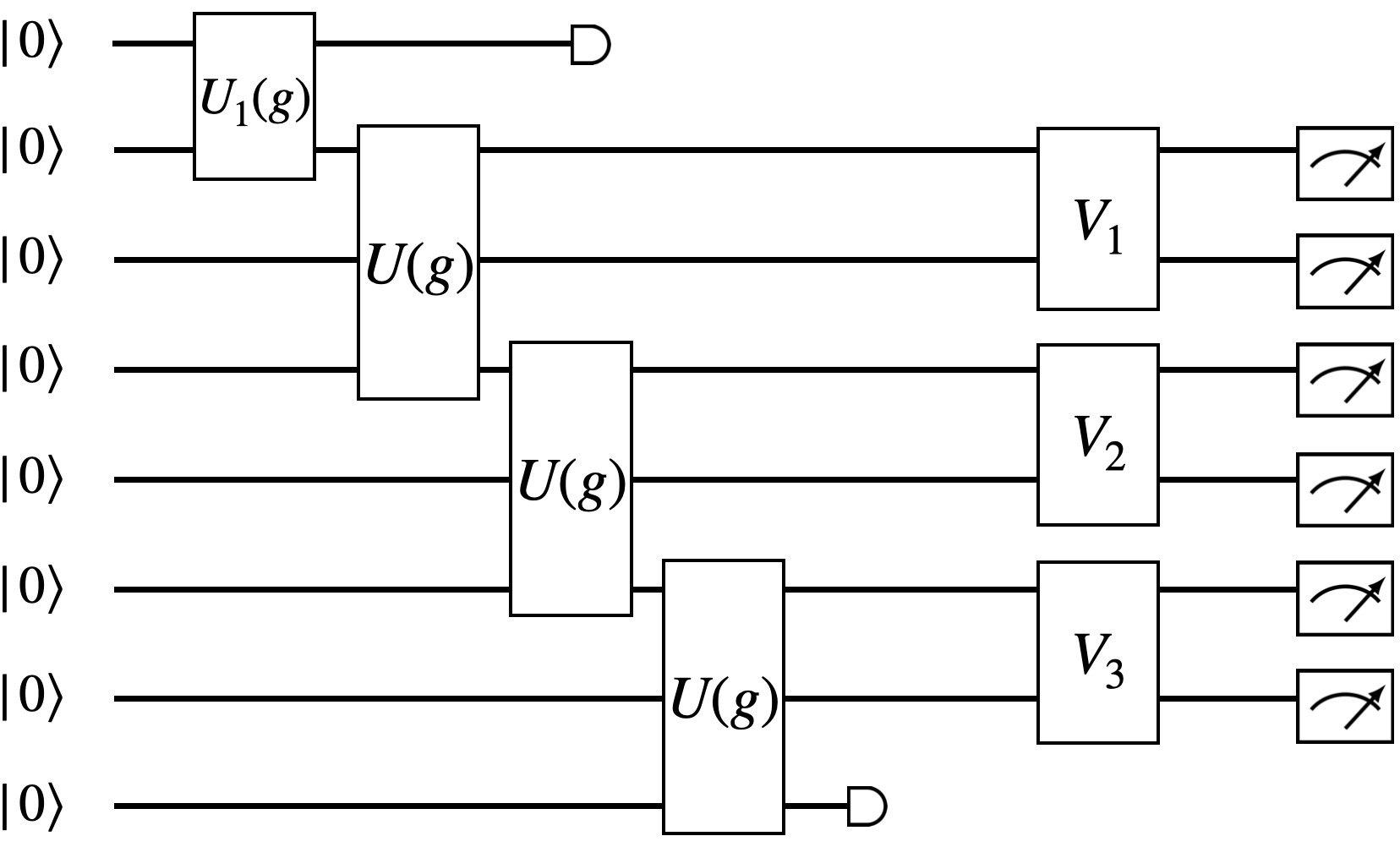}}\\[0.5cm]
\subfloat[Spin-$1$ family  (qutrits)\label{subfig:CircuitAKLTqutrits}]
{\includegraphics[width=0.8\columnwidth]{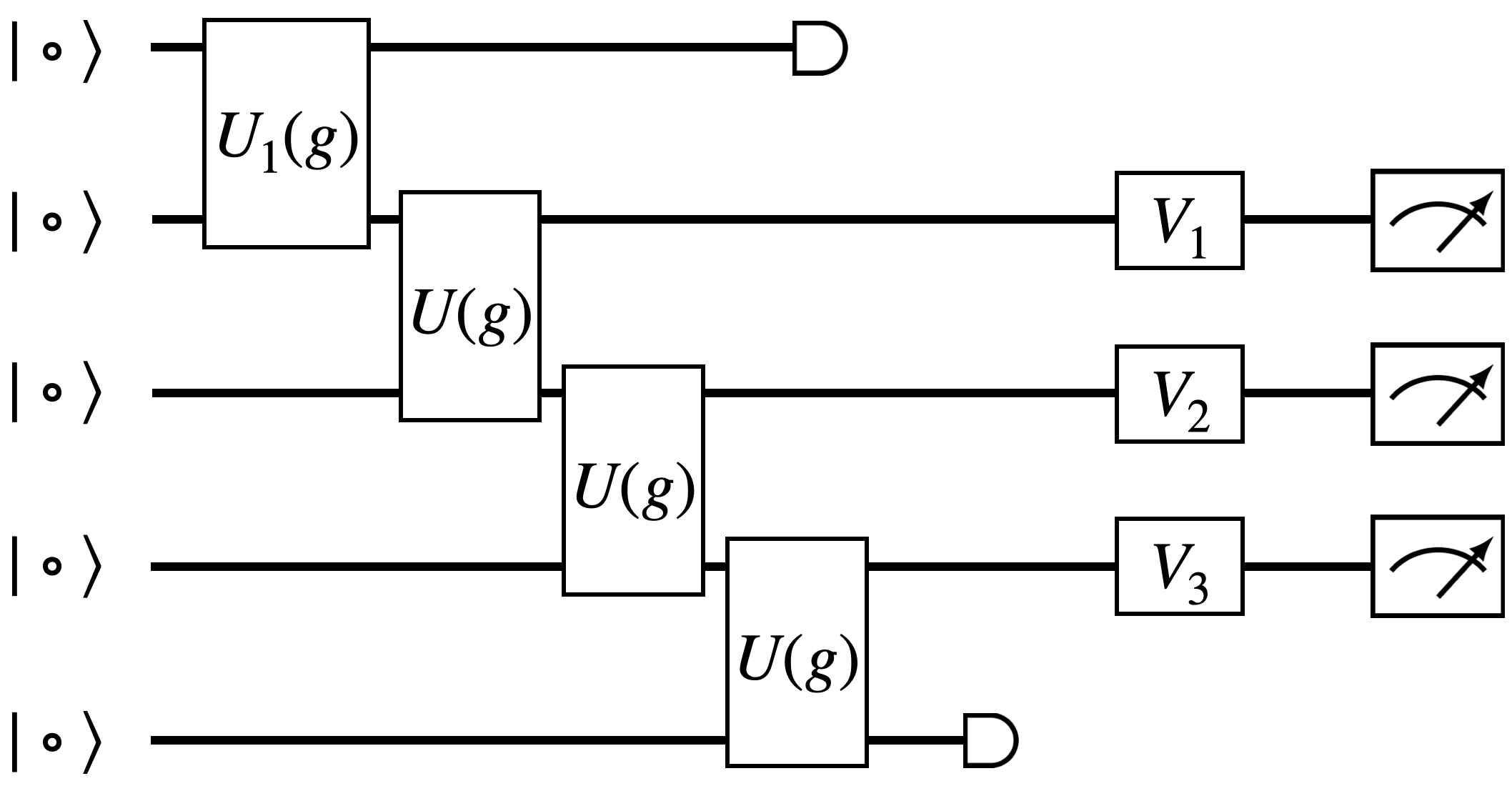}}
\caption{\textbf{Quantum circuits with randomized measurements}. Infinite-size MPS are represented as finite-size quantum circuits at the cost of two ancillary qubits or qutrits that are not measured. Each single-body unitary $V_{i}$ is drawn randomly with probability $1/(d+1)$ and rotates the physical degree of freedom into one of the MUB. Therefore, each configuration of the unitaries $V$ represents a distinct quantum circuit.}
\label{fig:QuantumCircuits}
\end{figure}

This ``unitarization'' is not unique, and it can be optimized to minimize the number of native gates necessary to implement the unitary matrix on the quantum hardware. 
Several algorithms solving this optimization problem exist and have been implemented as parts of open-source software tools such as \textit{UniversalQCompiler} \cite{Iten21} and \textit{BQSKit} \cite{Younis21}, which were both used in this work. 
To mimic an infinite system, the uppermost unitaries $U_1(g)$ in Fig.~\ref{fig:QuantumCircuits} are constructed from the non-trivial left eigenvector of the MPS transfer matrix and an analogous unitarization procedure is performed~\cite{Barratt21}. 

For the spin-$1/2$ family of states, a quantum circuit realization based on qubits is optimal, since the physical dimension of the MPS tensor equals its bond dimension. Therefore, an extra dummy leg of dimension $2$ turns the MPS tensor into a two-qubit gate that can be unitarized as explained above (cf. Fig.~\ref{fig:Unitarization}a).
This is not the case for the spin-$1$ states in Eq.~\eqref{eq:qutrit-mps}, for which physical and virtual dimensions are different. 
As a consequence, there are two possible ways to implement the spin-$1$ states on a quantum computer, utilizing either qubits or qutrits as computation units. 
On a qubit platform, the physical Hilbert space of a single spin-$1$ is emulated by a pair of qubits. The basis states $\ket{00},\ket{01},\ket{10}$ are associated with the spin-$1$ basis states $\ket{\circ},\ket{+},\ket{-}$, respectively, while the extra basis state $\ket{11}$ is neglected. The MPS tensor is then transformed into a 3-qubit gate by adding two dummy legs of dimension 2 (cf. Fig.~\ref{fig:Unitarization}b).
On a qutrit platform, on the other hand, the physical dimension of the MPS tensor matches the dimension of a qutrit, but the virtual Hilbert space dimension does not. Therefore the virtual Hilbert space is embedded into the subspace spanned by the qutrit basis states $\ket{+},\ket{-}$, and the state $\ket{\circ}$ is discarded. A 2-qutrit gate is then obtained by adding a dummy leg of dimension 3 (cf. Fig.~\ref{fig:Unitarization}c).


\section{Measurement}\label{sec:measurement}
We implemented informationally complete positive operator-valued measure (POVM) measurements (cf.~\cite{Carrasquilla2019}) based on mutually unbiased bases (MUB)~\cite{Wootters89}. The term mutually unbiased refers to the relation between the different orthogonal bases used for the POVM. Two orthogonal bases $\{ \ket{v_k} \}$ and $\{\ket{w_l}\}$ in a $d$-dimensional Hilbert space are mutually unbiased if the inner product of any pair of basis states always has the same magnitude: $\lvert \braket{v_k|w_l} \rvert = 1/\sqrt{d}$ for all $k,l$. 
When $d$ is the power of a prime, there are exactly $d+1$ MUB, and explicit constructions are known~\cite{Wootters89, Klappenecker05}. 
In such a case, a maximal set of $d+1$ MUB defines an informationally complete POVM via the collection $\{ M_l \}_{l=1}^{(d+1) d}$ of the projectors onto the MUB vectors. 

The best-known example of MUB is the set of eigenstates of the Pauli matrices
\begin{equation}\label{eq:MUBstates_d=2}
\resizebox{0.48\textwidth}{!}{$
\left\{ 
\begin{pmatrix}
1\\
0
\end{pmatrix},
\begin{pmatrix}
0\\
1
\end{pmatrix} 
\right\} \!,
 \frac{1}{\sqrt{2}} \! \left\{ 
 \begin{pmatrix} 
 1 \\
 1
 \end{pmatrix},
 \begin{pmatrix}
 1\\
 -1
 \end{pmatrix} 
 \right\} \!, 
 \frac{1}{\sqrt{2}} \! \left\{ 
\begin{pmatrix}
1 \\
i
\end{pmatrix},
\begin{pmatrix}
1\\
-i
\end{pmatrix}
\right\} \!,
$}
\end{equation}
that defines the Pauli-6 POVM, which we employed to take snapshots of the spin-$1/2$ family of states.

For the spin-$1$ case ($d=3$), we follow the construction of Ref.~\cite{Wootters89} and use the following maximal set of MUB 
\begin{equation}\label{eq:MUBstates_d=3}
\resizebox{0.48\textwidth}{!}{$
\begin{split}
&\left\{\begin{pmatrix}1\\0\\0\end{pmatrix},
	     \begin{pmatrix}0\\1\\0\end{pmatrix},
	     \begin{pmatrix}0\\0\\1\end{pmatrix}
\right\} \!,
\frac{1}{\sqrt{3}} \! \left\{
		 \begin{pmatrix}q_-\\1\\1\end{pmatrix},
		 \begin{pmatrix}1\\q_-\\1\end{pmatrix},
		 \begin{pmatrix}q_+\\q_+\\1\end{pmatrix}
\right\} \!, \\[0.2cm]
&\frac{1}{\sqrt{3}} \! \left\{
		 \begin{pmatrix}q_+\\1\\1\end{pmatrix},
		 \begin{pmatrix}1\\q_+\\1\end{pmatrix},
		 \begin{pmatrix}q_-\\q_-\\1\end{pmatrix}
\right\} \!,
\frac{1}{\sqrt{3}}\! \left\{
		 \begin{pmatrix}q_+\\q_-\\1\end{pmatrix},
		 \begin{pmatrix}q_-\\q_+\\1\end{pmatrix},
		 \begin{pmatrix}1\\1\\1\end{pmatrix}
\right\} \!,
\end{split}
$}
\end{equation}
where $q_\pm=e^{\pm 2\pi i/3}$. 

To implement these POVM measurements in the experiment we select a MUB with uniform probability $1/(d+1)$ for each qubit or qutrit, we rotate the latter with a single-site unitary $V_i$ into the selected MUB, and perform a measurement in the computational basis. The circuits realized in our experiments are depicted in Fig~\ref{fig:QuantumCircuits}, and discussed in more detail in Appendix~\ref{app:CircuitDetails}.

\section{Experimental setup}\label{sec:experimental_setup}
The experiment is performed on two trapped-ion devices. Both devices operate with $^{40}$\textrm{Ca}$^+$\ ions confined in macroscopic, linear Paul (`blade') traps. Information is encoded in Zeeman sublevels of $\textrm{4\,S}_{1/2}$ and $\textrm{3\,D}_{5/2}$ of $^{40}$\textrm{Ca}$^+$ ions which are connected via an optical quadrupole transition at \SI{729}{\nano\meter}. The optical systems of the respective traps provide single-ion addressing capabilities with \SI{729}{\nano\meter} light allowing for arbitrary single-qubit and arbitrary-pair two-qubit gate operations. The native gate set of the setups consists of single-qubit rotations around an axis in the equatorial plane of the Bloch sphere $R(\theta, \phi)$ implemented via laser pulses resonant with the optical transition, single-qubits $Z$-rotations ``virtually'' implemented in software, and maximally-entangling two-qubit gates $XX(\pm\pi/2) = \mathrm{exp}(\mp i\frac{\pi}{4} X\otimes X)$ implemented via the \text{M\o{}lmer-S\o{}rensen} interaction~\cite{sorensen1999}. For qubits, the $XX(\pm\pi/2)$ is equivalent to the CNOT gate up to local rotations~\cite{maslov2017}.

While the newer of the setups has the benefit of hardware upgrades enabling overall higher-performing operations, the second utilizes qutrits to its advantage. 

\paragraph{Qubit setup}
For the qubit experiment, the states are encoded in the magnetic Zeeman sublevels as \mbox{$\ket{0}=\ket{\textrm{4\,S}_{1/2},m_J = -1/2}$} and \mbox{$\ket{1}=\ket{\textrm{3\,D}_{5/2},m_J = -1/2}$}.
A detailed setup description can be found in Refs.~\cite{pogorelov2021, postler2022}.
Measurements are submitted as Qiskit~\cite{qiskit2024} circuits that undergo a custom transpilation procedure, the details of which are given in Appendix~\ref{app:qiskit}.

\paragraph{Qutrit setup}
For the qutrit experiment, the states are encoded in the magnetic Zeeman sublevels as \mbox{$\ket{+}=\ket{\textrm{4\,S}_{1/2},m_J = -3/2}$}, \mbox{$\ket{\circ}=\ket{\textrm{4\,S}_{1/2},m_J = -1/2}$}, and \mbox{$\ket{-}=\ket{\textrm{3\,D}_{5/2},m_J = -1/2}$}. The experimental setup used for the qutrit measurements~\cite{Edmunds:2024} has lower overall gate fidelities and shorter coherence times compared to the system used for the qubit experiments~\cite{pogorelov2021}. Due to geometric constraints, the optical single-ion addressing also results in larger crosstalk between ions and lower coupling strengths to the resonant transitions. Typical gate fidelities can be found in Ref.~\cite{Ringbauer2022}. The qutrit circuits were created using the compiler infrastructure in the python package BQSKit~\cite{BQSKit}.

\section{Machine Learning with tensorial-kernel SVM}\label{sec:ML}
Discriminating two physical phases of a many-body system can be viewed as a binary classification problem. For this purpose, supervised learning methods are frequently employed~\cite{Carrasquilla2017}, where a batch of training samples is collected, labelled according to their phase, and used to train the learning model. In supervised learning, the phase of the training samples is known beforehand. A much more challenging machine learning task, however, is determining the phase diagram and the features of each phase without any prior knowledge, \ie in an unsupervised fashion. This is precisely what the machine learning model presented here achieves.

Since any multi-classification task can be broken down into a set of binary classification tasks, we illustrate our method by considering a binary classification task. This implies that the training set consists of measurements sampled from only two different phase points (\eg different values of $g$ in Eq.~\eqref{eq:cluster-mps}). In our SVM-based approach, the two datasets are assigned distinct labels without knowing whether they belong to different physical phases or not. 
Rather than the user informing the machine about the relation between the two training sets, the machine informs the user via one of the model's internal parameters called the bias parameter. 

Crucially, the way in which the machine learns whether two datasets belong to different phases or the same phase is interpretable, meaning that the learning model contains information on local physical observables that is easy to retrieve from its parameters. 
The language in which this information is encoded in the learning model is specified by a local operator basis $\mathcal{B}$. Therefore, it is important to pick a minimal basis that captures all the non-trivial correlations.
The choice of the spin-$1/2$ operator basis is straightforward, and consists of the Pauli matrices: $\mathcal{B} = \{X,Y,Z\}$. For the spin-$1$ case, instead, we need to expand the basis of the $\tau$ matrices in Eq.~\eqref{eq:qutrit-paulis} to include their squares: $\mathcal{B} = \{\tau^x, \tau^y, \tau^z, (\tau^x)^2, (\tau^y)^2, (\tau^z)^2\}$. In principle, there are other non-trivial single-site matrices that can be included. However, as we show below, this minimal basis is enough to determine the correct phase diagram for the spin-$1$ MPS defined by the tensors in Eq.~\eqref{eq:qutrit-mps}.

Once the operator basis is specified, the first step of our algorithm is the computation of feature vectors $\bm{\phi}$ from the raw MUB samples. Their components comprise estimators of all possible monomials of a given rank $r$ on a given cluster of $n$ sites
\begin{equation}\label{eq:featureVector}
\bm{\phi}= \left( \left\langle O^{a_1}_{j_1} O^{a_2}_{j_2} \cdots  O^{a_r}_{j_r} \right\rangle \right)\, .
\end{equation}
Here, the multi-index $(a_1,\ldots,a_r; j_1,\ldots,j_r)$ labels the components of the feature vector: the indices $j_1,\ldots,j_r$ specify the sites within the predetermined cluster, while $a_1,\ldots,a_r$ label the operators chosen from the basis $\mathcal{B} = \{O^a\}$ acting on those sites.
%oact, while $a_1,\dots,a_r$ label the operators in the basis $\mathcal{B}=\{O^a\}$. 
The size $n$ of the cluster as well as the rank $r$ of the monomials, with $r \leq n$, are hyper-parameters. To eliminate redundancies, an ordering $j_1 < j_2 < \dots < j_r$ is imposed on the cluster indices, reducing the dimension of the feature vector to
\begin{equation}
\dim(\bm{\phi})=\binom{n}{r}\abs{\mathcal{B}}^r = \frac{n! \abs{\mathcal{B}}^r}{r!(n-r)!}.
\end{equation}
For example, the components of the feature vector with $r=2$ and $n=3$ for the spin-$1/2$ Pauli basis $\mathcal{B}=\{X,Y,Z\}$ are
\begin{equation}
\bm{\phi}=\left( \expect{X_1 X_2}, \expect{X_1 Y_2}, \expect{X_1 Z_2}, \dots, \expect{Z_2 Z_3} \right)
\end{equation}
and its dimension $\dim(\bm{\phi})=27$. Since no knowledge about the physical model besides the lattice geometry is assumed, it is in general not clear a priori how to choose the rank and cluster size. Therefore, we start with the lowest values of $r$ and $n$ and systematically increase them until convergence.

\begin{figure}
\centering
\subfloat[Separable \label{subfig:SVMproblemSeparable}]
{\includegraphics[width=0.9\columnwidth]{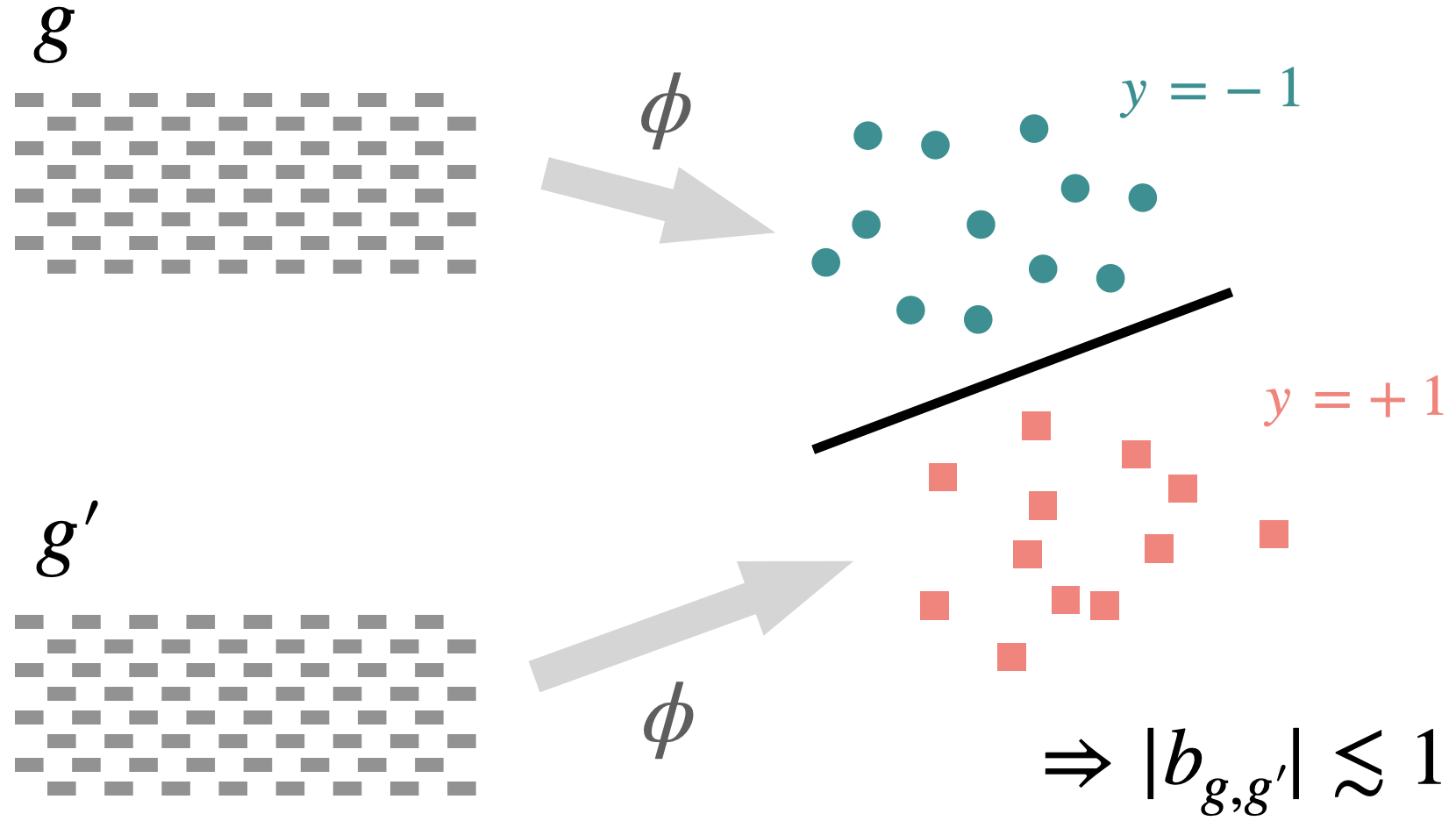}}\\[0.5cm]
\subfloat[Inseparable \label{subfig:SVMproblemNonSeparable}]
{\includegraphics[width=0.9\columnwidth]{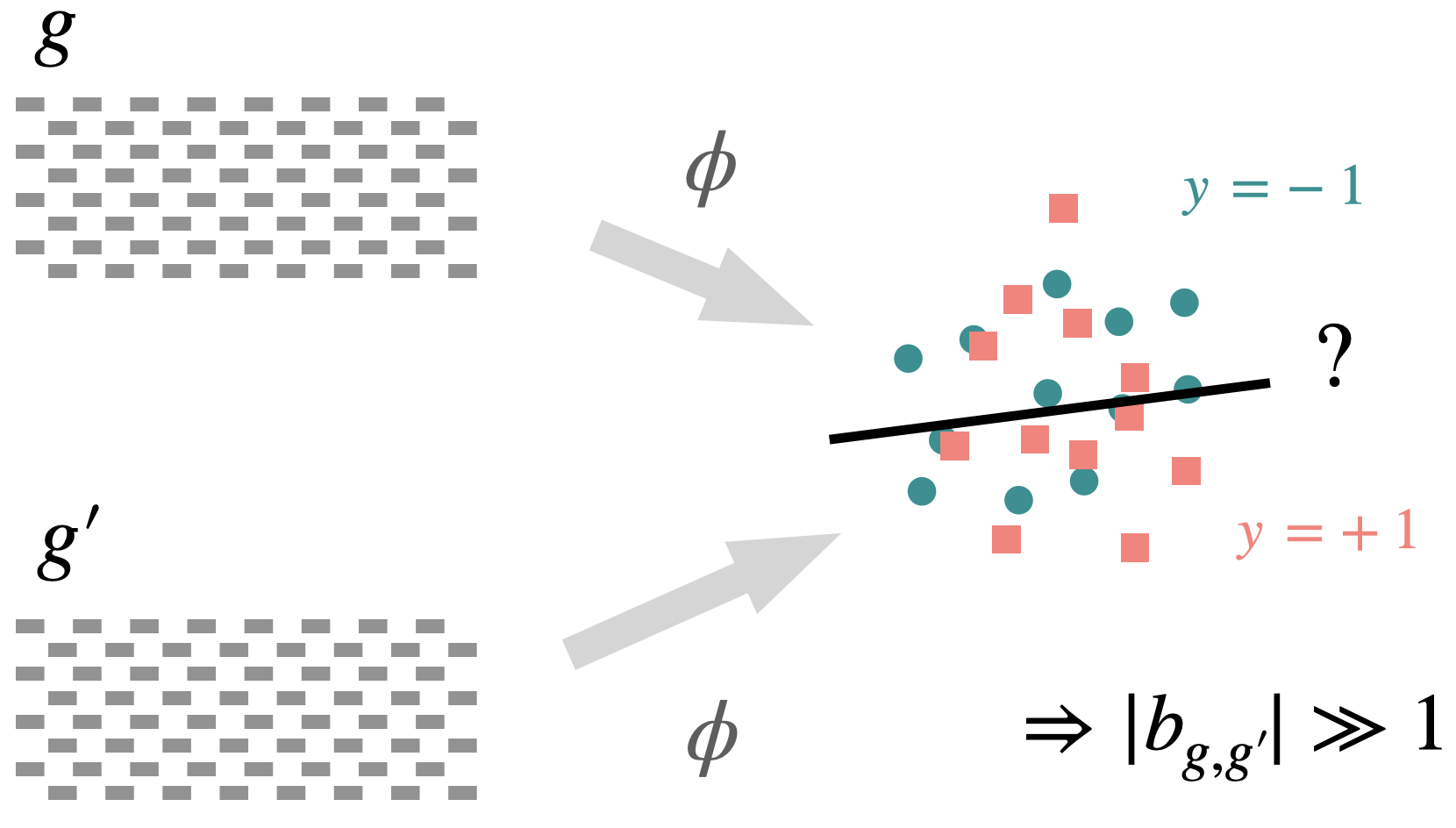}}
\caption{\textbf{Binary classification problem}. Two datasets of MUB samples labeled $g$ and $g'$ are mapped to two sets of feature vectors with class labels $y=-1$ and $y=+1$, respectively. (a) If $g$ and $g'$ are in different phases, the SVM successfully determines the hyperplane separating the classes in feature space, indicated by the absolute bias being close to unity. The decision function encodes the underlying order parameter. (b) If $g$ and $g'$ are in the same phase, the sets of feature vectors become inseparable and the absolute bias drastically exceeds unity. Because of this inseparability, the hyperplane has no physical meaning.}
\label{fig:SVMproblem}
\end{figure}
The components of the feature vector are statistical estimators of local observables, computed from the raw MUB samples by means of shadow tomography~\cite{Aaronson2018,Aaronson2019,Huang:2020}. Consider a set of $N_s$ independent MUB samples $\{M_1^{(l)},M_2^{(l)},\dots,M_N^{(l)}\}_{l=1}^{N_s}$ of a $N$-site state $\rho$, where each outcome $M^{(l)}_j$ is a projector onto one of the states from \eqref{eq:MUBstates_d=2} or \eqref{eq:MUBstates_d=3}, acting on site $j$. A single component of the feature vector in Eq.~\eqref{eq:featureVector} is then computed as
\begin{align}
& \left\langle O^{a_1}_{j_1} O^{a_2}_{j_2} \cdots  O^{a_r}_{j_r} \right\rangle =
\frac{1}{N_s} \sum_{l=1}^{N_s}
\mathrm{Tr} \big[O_{j_1}^{a_1} Q^{-1}(M^{(l)}_{j_1})\big] \times \nonumber \\ &  ~  \times
\mathrm{Tr} \big[O_{j_2}^{a_2} Q^{-1}(M^{(l)}_{j_2})\big] \times \cdots \times
\mathrm{Tr} \big[O_{j_r}^{a_r} Q^{-1}(M^{(l)}_{j_r})\big],
\end{align}
with $Q^{-1}(A) = (d+1)A - \mathrm{Tr} (A)I$ the inverted single-site quantum channel~\cite{Huang:2020} fixed by the POVM. More details on shadow tomography can be found in Appendix~\ref{app:ClassicalShadow}.

For each value of $g$, we construct $N_s$ feature vectors 
$\{\bm{\phi}^{(k)}\}_{k=1}^{N_s}$
and apply our TK-SVM method to determine whether the datasets $\{\bm{\phi}_g^{(k)}\}$ and $\{\bm{\phi}_{g'}^{(k)}\}$, sampled at different MPS parameters $g$ and $g'$, belong to the same phase or not. To this end, we assign class labels $y_k = \pm 1$ to the two sets of feature vectors (e.g., $y_k = +1$ for samples at $g$ and $y_k = -1$ for samples at $g'$).
These two labeled classes are then provided as input to the SVM, whose optimization objective is to determine a separating hyperplane between the datasets~\cite{Boser:1992}. If $g$ and $g'$ correspond to different phases, the two classes of feature vectors become separable and the SVM successfully identifies a separating hyperplane (see Fig.~\ref{subfig:SVMproblemSeparable}). Conversely, if $g$ and $g'$ belong to the same phase, the classes are inseparable and the SVM fails to find a separating solution (see Fig.~\ref{subfig:SVMproblemNonSeparable}). 

Regardless of separability, the SVM returns the optimal parameters defining a decision function $D(\bm{\phi})$, which maps any test sample $\bm{\phi}$ to a real number. The sign of $D(\bm{\phi})$ determines the predicted class of $\bm{\phi}$, \ie on which side of the separating hyperplane the sample lies. In this work, we employ a linear-kernel SVM, whose decision function reads
\begin{equation} \label{eq:decisionFunction}
D(\bm{\phi}) = \sum_k \lambda_k y_k  K( \bm{\phi}^{(k)},\bm{\phi} ) - b,
\end{equation}
where the kernel is given by $K( \bm{\phi}^{(k)},\bm{\phi} ) = \bm{\phi}^{(k)} \cdot \bm{\phi}$. The coefficients $\lambda_k$ and the bias $b$ are internal SVM parameters whose optimal values depend on the datasets at $g$ and $g'$. The index $k$ in \cref{eq:decisionFunction} runs over the combined datasets at $g$ and $g'$, i.e., $k = 1,2,\dots, 2N_s$.

Determining the separating hyperplane, \ie training the SVM to obtain the optimal parameters $\lambda_k$ and $b$, constitutes a quadratic optimization problem for which standard solution methods are available, with guaranteed polynomial scaling in the number of feature vectors~\cite{Nocedal2006}.

The inability to determine a separating hyperplane is reflected in the bias parameter $b$, which allows us to determine whether the two datasets belong to the same physical phase using the heuristic bias criterion put forward in Ref.~\cite{Greitemann_PhD:2019}
\begin{equation}
\abs{b_{g,g'}}
\begin{cases}
\lesssim 1, & g,g' \text{ in different phases},\\
\gg 1,  & g,g' \text{ in the same phase}.
\end{cases}
\end{equation}
Based on this criterion, the phase diagram can be obtained systematically. This is achieved by performing binary classification between pairs $(g,g')$ to get the bias parameter $b_{g,g'}$. The absolute values $|b_{g,g'}|$ are then used to weight~\footnote{The edge weight is obtained by normalizing the absolute bias using a Lorentzian function, as described in~\cite{sadoune23}.} the edges of the graph whose vertices are labeled by the distinct values of $g$. Since all points of the same phase are strongly connected (large absolute bias) and points of different phases are weakly connected, partitioning the graph directly yields the topology of the phase diagram. 

To obtain the graph partitions, i.e. the different quantum phases, we apply Fiedler theory~\cite{Fiedler73,Fiedler75}, which assigns a real value, the Fiedler value, to each vertex. These values are the components of the eigenvector associated with the second-smallest eigenvalue of the Laplacian matrix of the graph, the so-called Fiedler vector. The presence of disconnected graph components, i.e. the distinct quantum phases, becomes clear from the appearance of groups of sites with similar Fiedler values.
For a bipartite graph, the sign of the Fiedler value indicates which phase the vertex belongs to. 
\begin{figure}
\centering
\subfloat[Spin-$1/2$ family \label{subfig:PhasesCluster}]
{\includegraphics[width=1.\columnwidth]{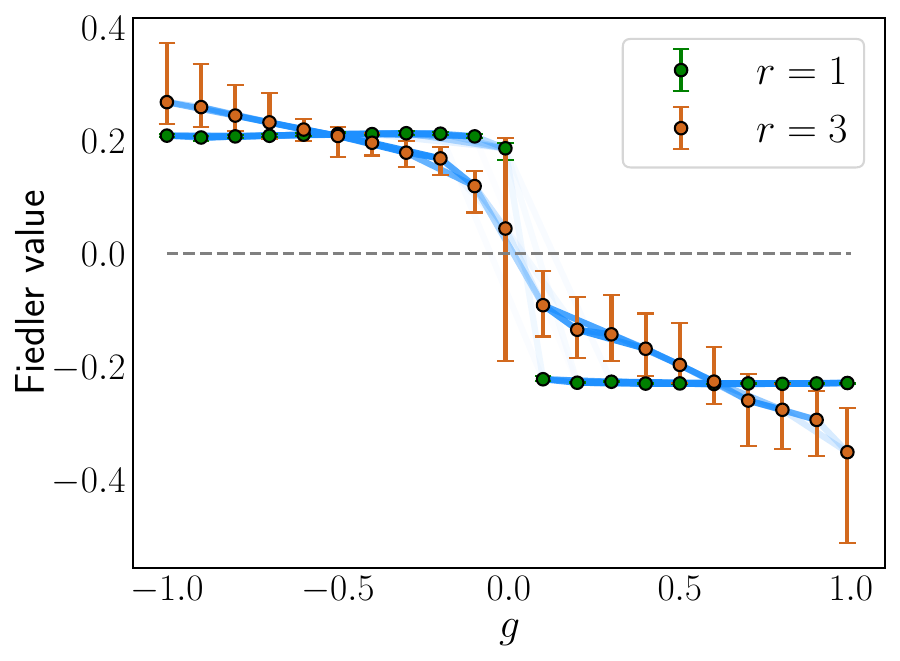}}\\
\subfloat[Spin-$1$ family (qubit platform)\label{subfig:PhasesAKLTqubits}]
{\includegraphics[width=1.\columnwidth]{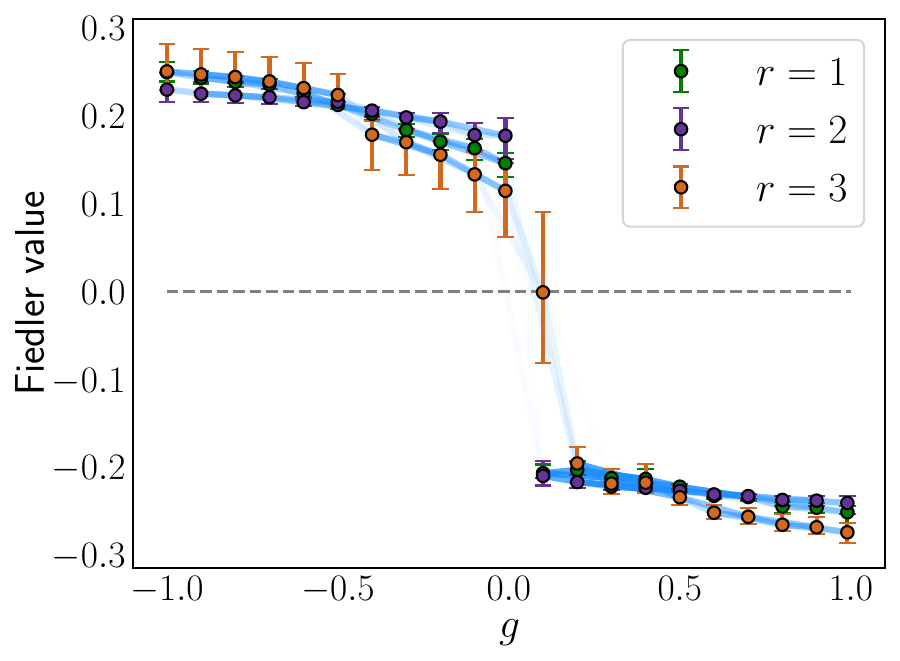}}
\caption{\textbf{TK-SVM phase classification.} Fiedler value of the weighted graphs constructed from the TK-SVM bias parameters, which appear to be bipartite graphs both in spin-$1/2$ (a) and spin-$1$ (b) cases, indicating the presence of two phases. The normalized weight of the edges is represented by their opacity. The sign of the Fiedler value indicates which component of the graph, \ie which phase, it belongs to. Error bars are obtained by Jackknife resampling~\cite{Efron81}. The phase classification result of the spin-$1$ family realized on the qutrit platform is displayed in Appendix~\ref{app:QubitQutritComparison}.}
\label{fig:PhaseClassification}
\end{figure}

Once the distinct phases are identified, all datasets in the same phase are pooled and TK-SVM is run once more, this time classifying each phase against random uniformly generated MUB samples, meaning that the class label $y_k = 1$ now refers to the datasets of an entire phase
and $y_k = -1$ refers to the random samples.
Upon training the SVM, the decision function Eq.~\eqref{eq:decisionFunction} encodes the physical quantity discriminating the two phases, \eg an order parameter. 
By expanding the inner product in Eq.~\eqref{eq:decisionFunction}, the decision function is re-expressed as
\begin{equation}
\begin{split}
D(\bm{\phi}) &= \sum_{\mu} C_{\mu} \phi_\mu - b, \\ 
C_{\mu} &= \sum_k \lambda_k y_k \phi_\mu^{(k)},
\end{split}
\end{equation}
where we introduced a composite index labeling the components of a feature vector $\bm{\phi}=\{\phi_\mu\}$. 

The resulting coefficient vector $C_{\mu}$ is constructed from the training data weighted by the SVM optimization parameters and thus contains all the desired information. 
Specifically, its non-vanishing components correspond to the physical features based on which the machine makes predictions. Note that the decision function is never used to predict the class of a test sample. Instead, we merely extract the learning model parameters to gain information on the relation between two phase points (from the bias $b_{g,g'}$) or on the characteristic features of a whole phase (from the coefficient vector $C_{\mu}$). 
In Appendix~\ref{app:Accuracy}, however, we will use TK-SVM to make predictions on a test set for benchmarking purposes.

\section{Phase classification and characterization results}\label{sec:results}
\begin{figure}
\centering
\subfloat[$g>0$ rank $1$\label{subfig:clusterModel_feat_T200_r1}]
{\includegraphics[width=0.9\columnwidth]{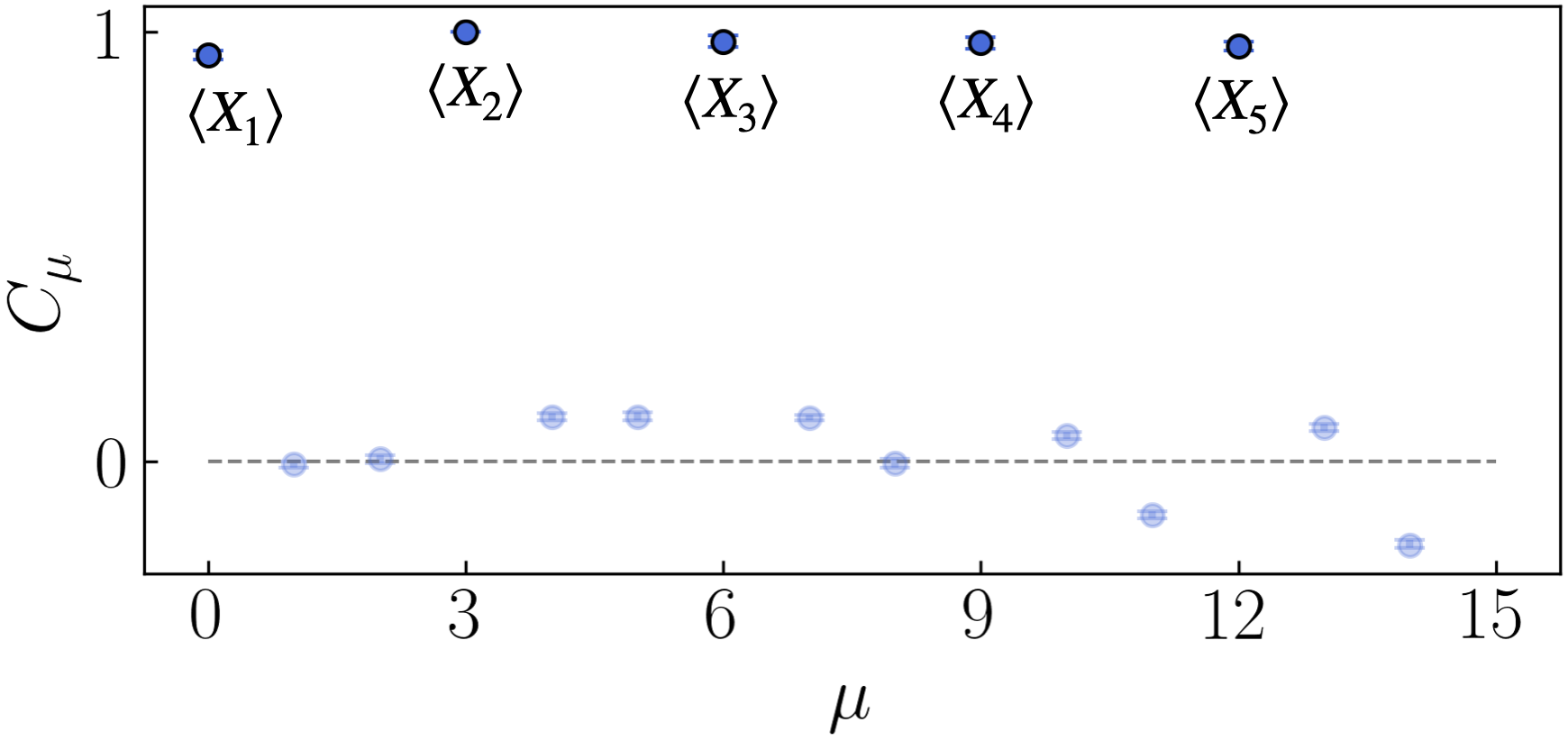}}\\
\subfloat[$g<0$ rank $3$\label{subfig:clusterModel_feat_T100_r3}]
{\includegraphics[width=0.9\columnwidth]{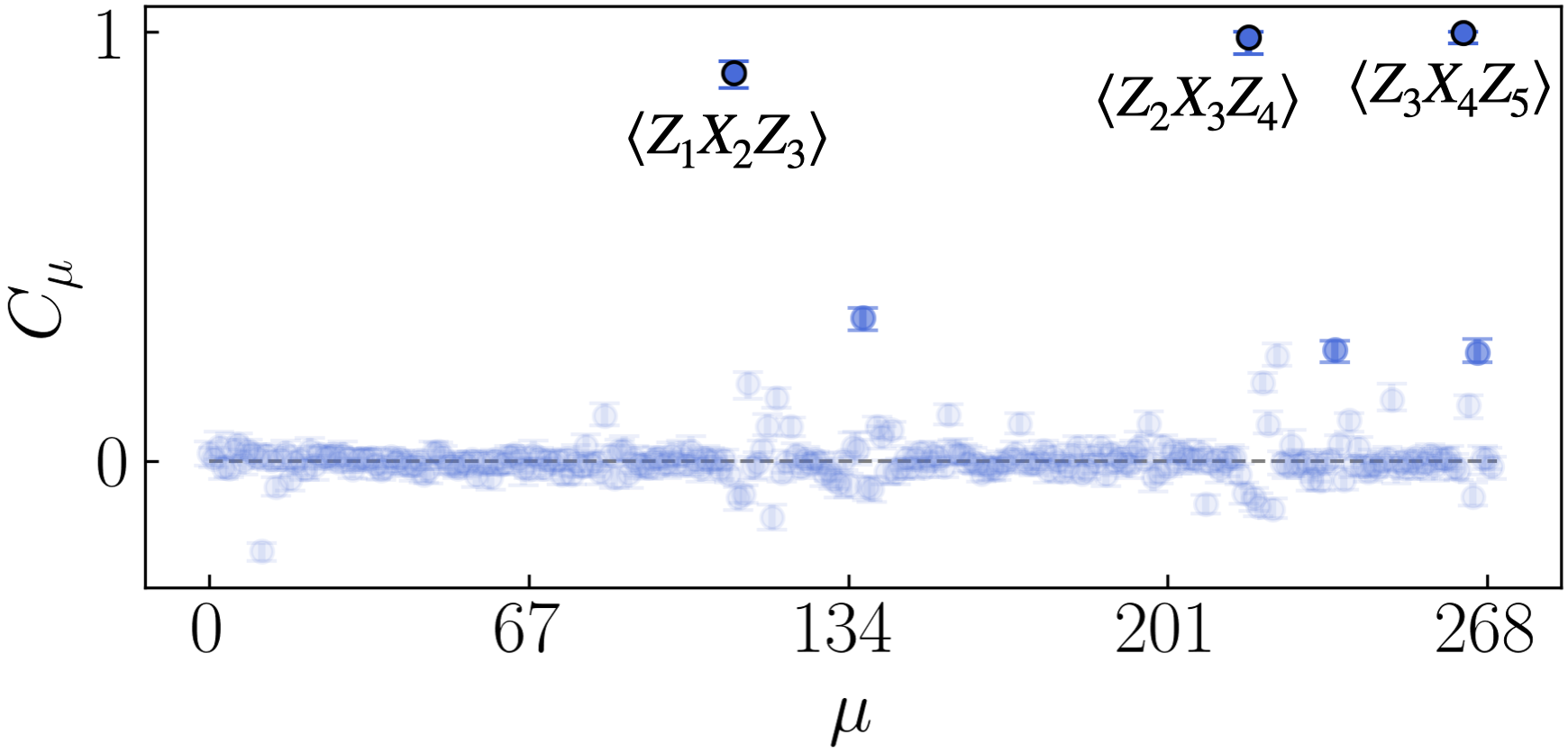}}\\
\subfloat[$g<0$ rank $4$\label{subfig:clusterModel_feat_T100_r4}]
{\includegraphics[width=0.9\columnwidth]{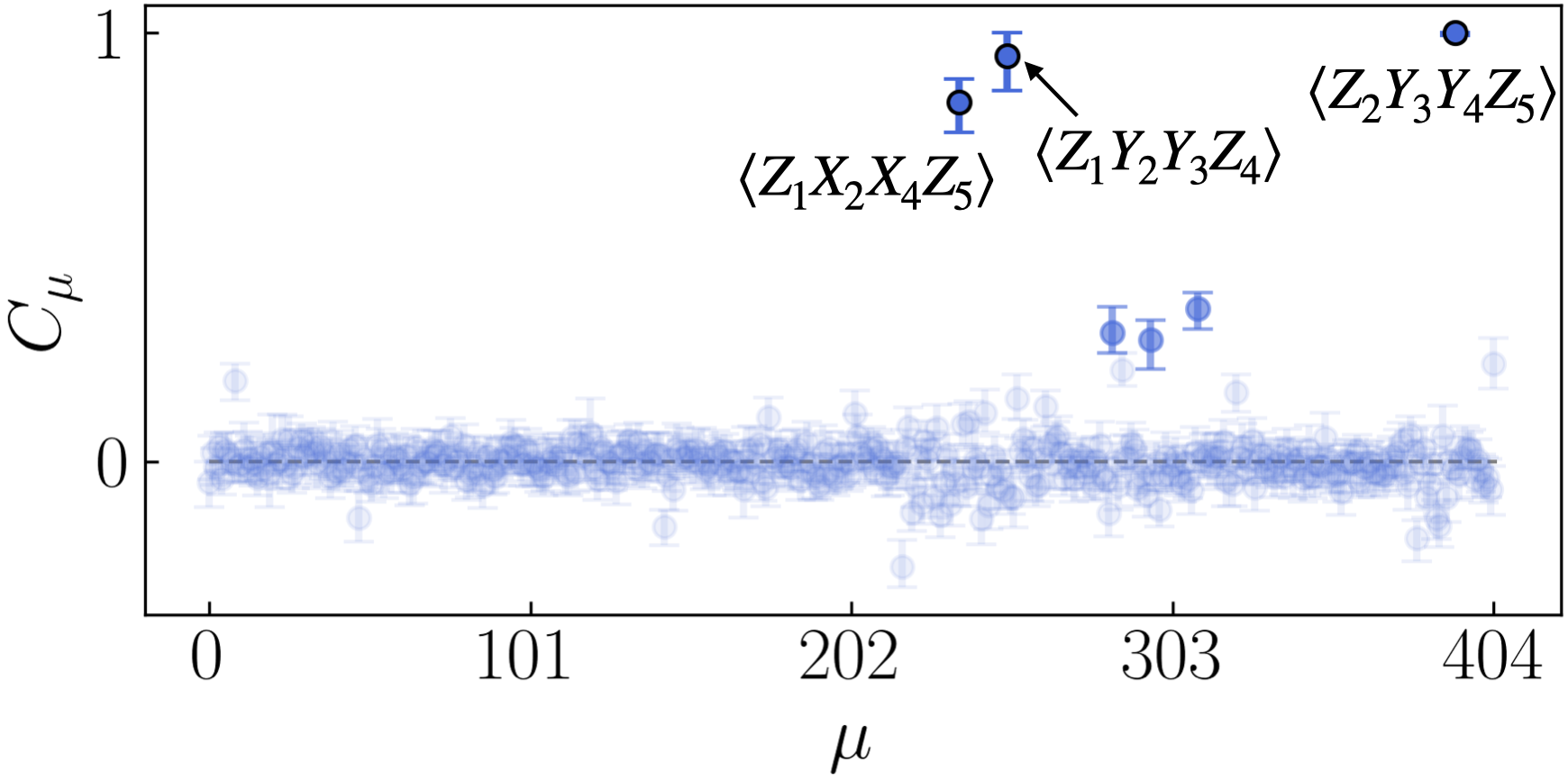}}\\
\subfloat[$g<0$ rank $5$\label{subfig:clusterModel_feat_T100_r5}]
{\includegraphics[width=0.9\columnwidth]{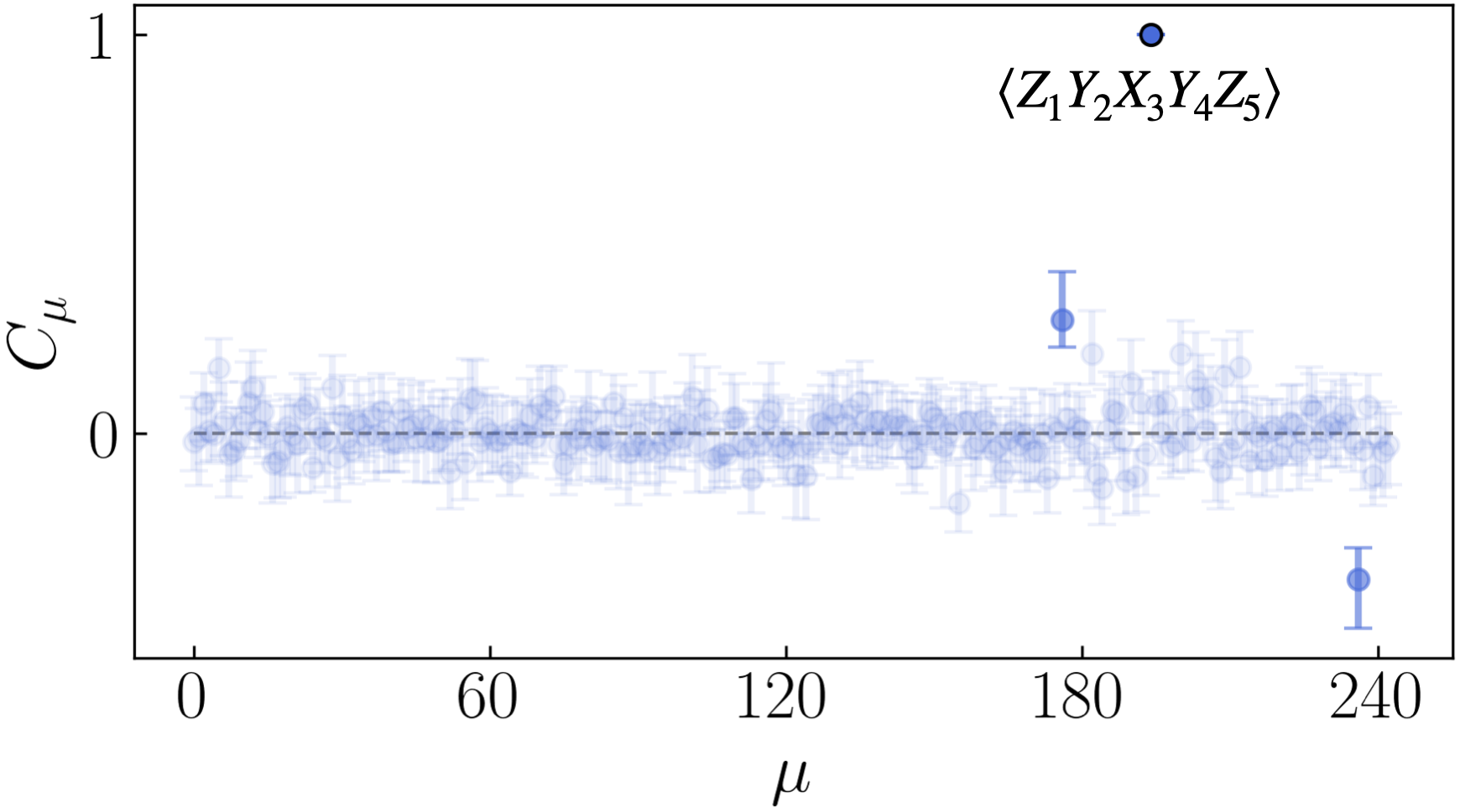}}
\caption{\textbf{Coefficient vector $C_{\mu}$ for the spin-$1/2$ family} for different merged datasets and ranks. The cluster size is fixed at $n=5$ in all cases, matching the whole system size. Each index $\mu$ is associated with an $r$-point correlation function. Entries with the largest absolute coefficients are labeled with their corresponding $r$-point expectation value.}
\label{fig:clusterModel_feat}
\end{figure}

For each family of states, we consider $21$ values of $g$ picked equidistantly in the interval $g\in [-1, 1]$. The spin-$1/2$ family of states is realized on 5 physical and 2 ancillary qubits, resulting in $3^5=243$ possible measurement configurations, \ie choices of the unitaries $V$ in Fig.~\ref{fig:QuantumCircuits}a. For each measurement configuration, $500$ projective samples are collected, i.e. $121500$ MUB samples per value of $g$. The spin-$1$ family of states, instead, is realized on 3 physical and 2 ancillary spin-1 sites with a physical local Hilbert space dimension $d=3$, implying $4^3=64$ possible measurement configurations for the unitaries $V$ in Fig.~\ref{fig:QuantumCircuits}b and Fig.~\ref{fig:QuantumCircuits}c. When such a state is realized on the qubit platform, where each spin-$1$ site is emulated by two qubits, we collect $2000$ projective samples per measurement configuration. Thus, the number of MUB samples totals $128000$ per value of $g$. Due to the imperfect realization of this state in the experiment, however, a fraction of the samples contains the projective qubit pair outcome $\ket{11}$. These samples have no valid spin-$1$ basis state correspondence and are thus discarded. The fraction of discarded samples ranges from $14\%$ to $22\%$ depending on $g$, resulting in an average of about $106000$ usable MUB samples for each value of $g$. 
From the qutrit platform, instead, we collect $3000$ samples per measurement configuration, none of which has to be discarded. This amounts to $192000$ MUB samples per value of $g$.
The resulting machine-learned phase diagrams are displayed in Fig.~\ref{fig:PhaseClassification}. For the spin-$1$ family of states, the qubit platform yields slightly better results regarding phase classification, and a comparison between qubit and qutrit platforms can be found in Appendix~\ref{app:QubitQutritComparison}.

\begin{figure}
\centering
\subfloat[$g>0$ rank 1\label{subfig:aklt-productstate_feat_T200_r1}]
{\includegraphics[width=0.9\columnwidth]{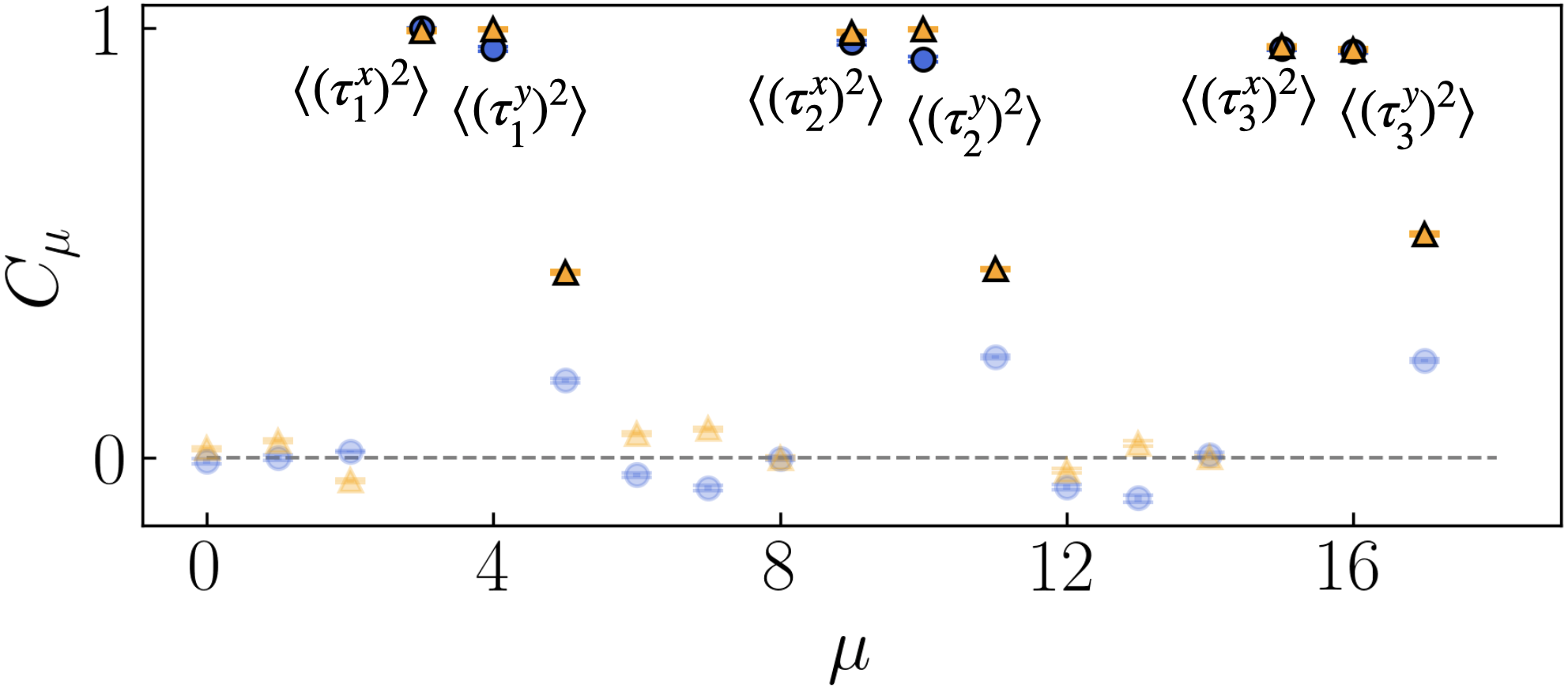}}\\
\subfloat[$g<0$ rank 2\label{subfig:aklt-productstate_feat_T100_r2}]
{\includegraphics[width=0.9\columnwidth]{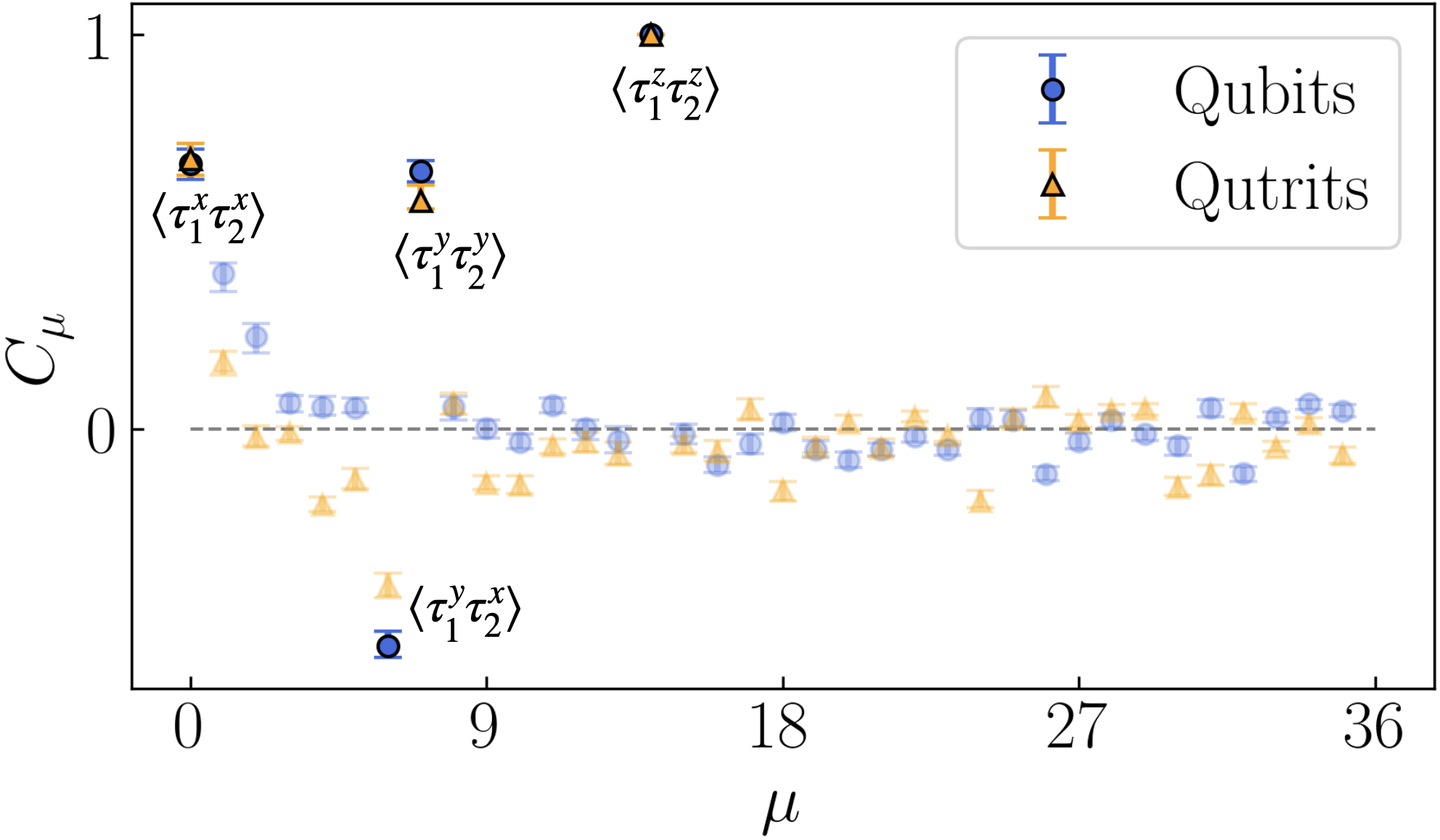}}\\
\subfloat[$g<0$ rank 3\label{subfig:aklt-productstate_feat_T100_r3}]
{\includegraphics[width=0.9\columnwidth]{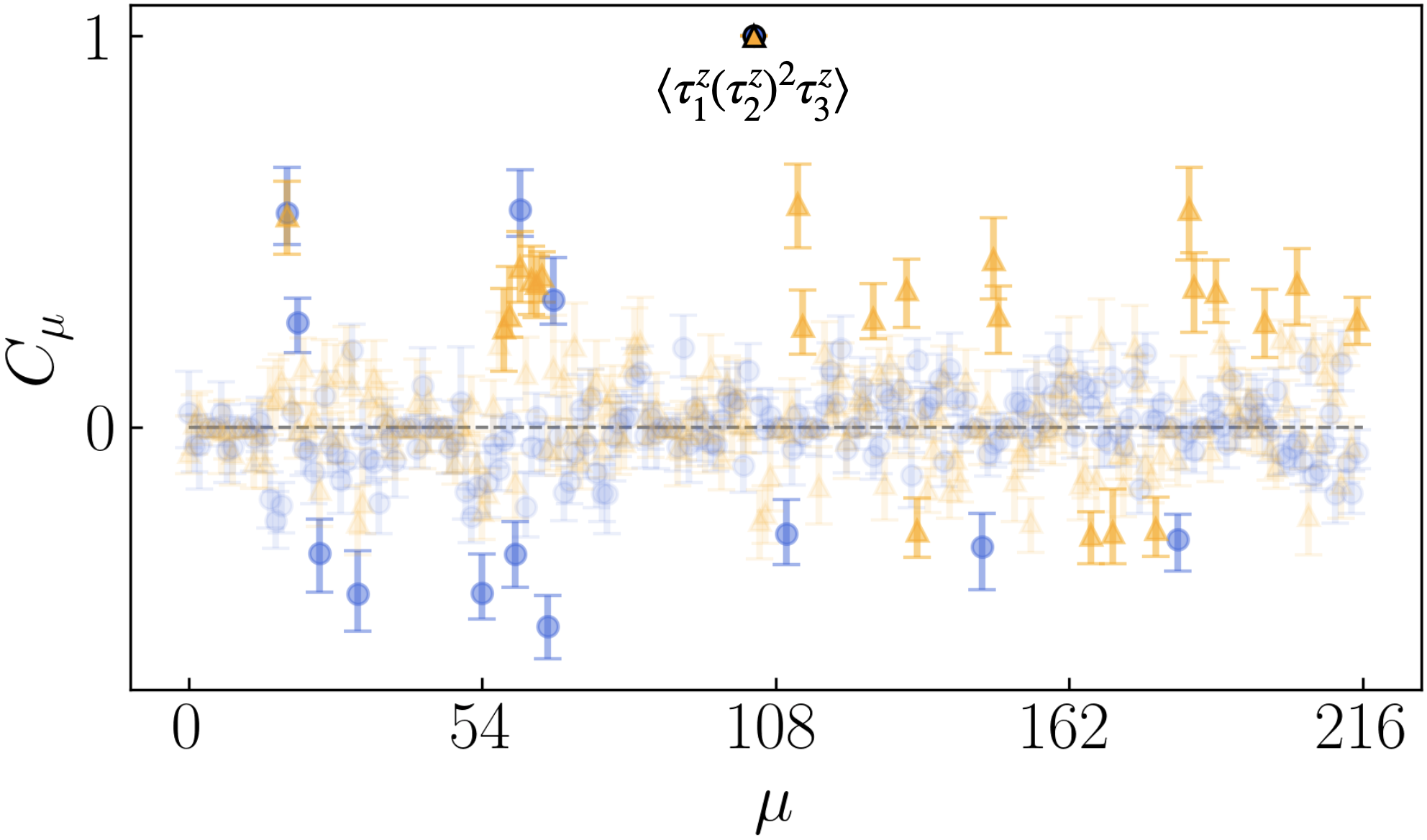}}
\caption{\textbf{Coefficient vector for the spin-$1$ family.} For (a) and (c) a cluster size $n=3$ is used while for (b) the cluster size is $n=2$. In all cases the outcomes of qubit and qutrit platform agree on the most dominant features. Only at rank $3$ there is a discrepancy regarding the sub-leading features. An important difference to the spin-$1/2$ family is that the operator basis for spin $1$ is able to produce the identity $(\tau^x)^2 + (\tau^y)^2 + (\tau^z)^2 = 2I$. The rank $2$ features $\expect{\tau^z_1 \tau^z_2}$ would thus appear in the rank $3$ result as the sum over three features $\sum_{a=x,y,z}\expect{\tau^z_1 \tau^z_2 (\tau^a_3)^2} = 2\expect{\tau^z_1 \tau^z_2 I_3}$. All redundant features of this sort are masked automatically to simplify interpretation.}
\label{fig:aklt-productstate_feat}
\end{figure}

As explained in Sec.~\ref{sec:ML}, we now proceed by pooling data corresponding to the same phase, thereby obtaining one dataset for $g < 0$ and one dataset for $g>0$, and classify them against a set of random uniform MUB samples that lack any kind of order at any rank.
This procedure has the advantage that subleading features that are not characteristic of the whole phase are suppressed. 
Extracting and interpreting the coefficient vector $C_{\mu}$ we get the order parameters distinguishing the two phases.
We show below that our phase classification algorithm correctly identifies the two different phases. 

We start with the spin-$1/2$ family. Its rank $r = 1$ coefficient vector for the $g>0$ pooled dataset (trivial paramagnetic phase) is displayed in Fig.~\ref{subfig:clusterModel_feat_T200_r1}. For the cluster of size $n=5$ we find the same prominent local observable $\expect{X}$ at each site. Hence, the minimal unit cell that captures the main feature of this state is a single site, and choosing the hyperparameter $n=1$ would have been sufficient. At higher ranks (not shown), only redundant products of the features already observed for $r = 1$ arise, such as $\expect{X_1 X_2}$, $\expect{X_3 X_4 X_5}$, and so on. Up to $r = 5$, no significant features appear that cannot be reduced to products of rank-$1$ features. Therefore we conclude that the underlying local order characterizing this phase is given by
\begin{equation}
\mathcal{O}_{g>0}=  X_j.
\end{equation}

For $g< 0$ (SPT phase), the rank-$1$ decision function has an absolute bias drastically exceeding unity, implying that TK-SVM detects no qualitative difference w.r.t. the random samples. Hence, the resulting coefficient vector has no meaning and it is not shown. The same holds for $r=2$. Increasing the rank further, the bias parameter approaches unity and distinct features become apparent in the coefficient vector, see Figs.~\ref{subfig:clusterModel_feat_T100_r3},~\ref{subfig:clusterModel_feat_T100_r4}, and~\ref{subfig:clusterModel_feat_T100_r5}. The higher rank features can all be understood as non-trivial overlapping products of the rank-$3$ features $G_j:=Z_j X_{j+1} Z_{j+2}$. For example, the rank-$5$ feature $\expect{Z_1 Y_2 X_3 Y_4 Z_5}$ is the overlapping product $\expect{G_1 G_2 G_3}$. The overlap can either span two sites as in the previous example or a single site as in $\expect{G_1 G_3}\propto\expect{Z_1 X_2 X_4 Z_5}$. Extrapolating these observations to higher ranks, we conclude that the underlying order is of the form
\begin{equation}
\mathcal{O}^2_{g<0} = \prod_{i=j}^{r-2} G_j \propto Z_1 Y_2 \bigg[\prod_{j=1}^{r-2} X_j\bigg] Y_{r-1} Z_{r},
\end{equation}
for two-site overlap, and
\begin{equation}
\mathcal{O}^1_{g<0} = \prod_{j=1}^{(r-1)/2} G_{2j-1} \propto Z_1 \bigg[\prod_{j=1}^{(r-1)/2} X_{2j}\bigg] Z_{r},
\end{equation}
for single-site overlap and assuming $r$ is odd. Both expressions are typically found in the literature as string order parameters.

We now discuss the spin-$1$ family of states. Analogous to the spin-$1/2$ family, only the phase $g>0$ (trivial phase) is distinguishable from the random samples at rank $1$, in which case we obtain the coefficient vector displayed in Fig.~\ref{subfig:aklt-productstate_feat_T200_r1}. The periodicity w.r.t. the three sites of the cluster implies that the underlying order extends merely over a single site. Hence, the order parameter of the trivial phase can be inferred to be
\begin{equation}
\mathcal{O}_{g>0} =  (\tau_j^x)^2 + (\tau_j^y)^2 .
\end{equation}
As expected for the trivial phase, the predominant features at higher ranks are always reducible to products of the rank $1$ features. This indicates the proximity to a product state throughout the phase, and the product state saturating the order parameter is precisely the state in the $g= 1$ limit: $\bigotimes \ket{\circ}$.

In the phase $g<0$ (SPT) we observe the dominant features $\expect{\tau_1^z \tau_2^z}$ at rank $2$ and $\expect{\tau_1^z (\tau_2^z)^2 \tau_3^z}$ at rank $3$, see Figs.~\ref{subfig:aklt-productstate_feat_T100_r2} and~\ref{subfig:aklt-productstate_feat_T100_r3}, respectively. We extrapolate these observations to infer the string order for the $g<0$ phase:
\begin{equation}
\mathcal{O}_{g<0}=\tau_1^z \bigg[ \prod_{j=2}^{(r-1)} (\tau_j^z)^2 \bigg] \tau_r^z.
\end{equation}
In contrast to the spin-$1/2$ case, the Hermitian operators of the spin-$1$ basis are not unitary. Therefore, it is crucial to normalize the string operator to have a non-vanishing string order for $r \to \infty$.
By setting $\mathcal{N} \int \mathrm{d} \psi  \, \langle \psi | (\tau^z)^2 | \psi \rangle  =1$, where the states $\psi$ are Haar-distributed, we get $\mathcal{N} = 3/2$.
With this normalization, the expectation value of the rescaled string operator matches the expectation value of the well-known AKLT string order parameter
\begin{multline}
\langle \psi(g) | \tau_1^z \bigg[ \prod_{j=2}^{(r-1)} \frac{3}{2}(\tau_j^z)^2 \bigg] \tau_r^z  | \psi(g) \rangle \\ =
(-1)^r \langle \psi(g) | \tau_1^z \bigg[ \prod_{j=2}^{(r-1)} e^{i\pi \tau^z_j} \bigg] \tau_r^z | \psi(g) \rangle ,
\label{eq:aklt-string}
\end{multline}
where we used that $e^{i\pi \tau^z}  = -3 (\tau^z)^2/2 + (\tau^x)^2/2 + (\tau^y)^2/2  $ and that the expectation value of all observables containing $(\tau^x)^2$ or $(\tau^y)^2$ vanish, when expanding the exponential in the r.h.s of Eq.~\eqref{eq:aklt-string}.

\section{Conclusions and Outlook} \label{sec:outlook}
We analyzed quantum data from state-of-the-art trapped-ion quantum computing platforms using unsupervised, interpretable machine learning based on tensorial-kernel support machines. 
In particular, we implemented two families of MPSs featuring a transition between a trivial phase and a symmetry-protected topological phase. 
The first family represents a well-studied spin-1/2 system that interpolates between the cluster state and a trivial product state, while the second is a less-explored spin-1 system with the AKLT state as a paradigmatic instance of SPT order.
We utilized informationally complete measurements obtained experimentally and demonstrated the robustness of TK-SVM to experimental imperfections, unavoidable in present-day quantum devices. 
We did so by showing that our method correctly distinguishes between the two phases in both setups and that the machine learning model itself is interpretable, allowing us to successfully extract the string order parameters characterizing the SPT phase.

Our results highlight the utility of classical machine learning for the efficient investigation of quantum phase diagrams in NISQ devices. 
Due to hardware constraints, we focused on quantum data from systems with a small number of qudits. However, our machine learning approach is scalable and can be applied to larger quantum systems without modification, opening the door to practical quantum advantage even without error correction. In fact, a theoretical analysis using synthetic data shows that the TK-SVM method scales well in accuracy as both the number of qudits and the required sample size increase (see Appendix~\ref{app:Accuracy}).

Natural extensions of this work include applying our method to the adiabatic state preparation typical of analog quantum simulators. 
There, the challenge is to ensure informationally complete measurements, for which significant progress in the case of Rydberg atoms was recently reported~\cite{bornet2024enhancing}. 
%In this setting, TK-SVM would enlarge the toolbox of available techniques to analyze large-size quantum systems in an unbiased fashion, with potential implications for the solution of long-standing problems not amenable to classical algorithms for many-body systems.

\begin{acknowledgments}
We acknowledge useful discussions with Ke Liu.
N.S. and L.P. acknowledge support from FP7/ERC Consolidator Grant QSIMCORR, No. 771891, and the Deutsche Forschungsgemeinschaft (DFG, German Research Foundation) under Germany’s Excellence Strategy -- EXC-2111 -- 390814868.
Giuliano Giudici acknowledges support from the European Union’s Horizon Europe program under the Marie Sk\l{}odowska Curie Action TOPORYD (Grant No. 101106005).
The project/research is part of the Munich Quantum Valley, which is supported by the Bavarian state government with funds from the Hightech Agenda Bayern Plus.
The Innsbruck team acknowledges support by the European Research Council (ERC, QUDITS, 101039522), and the European Union’s Horizon Europe research and innovation programme under grant agreement No 101114305 (``MILLENION-SGA1'' EU Project). Views and opinions expressed are however those of the author(s) only and do not necessarily reflect those of the European Union or the European Research Council Executive Agency. Neither the European Union nor the granting authority can be held responsible for them. We also acknowledge support by the Austrian Science Fund (FWF Grant-DOI 10.55776/F71) (SFB BeyondC), the Austrian Research Promotion Agency under Contracts Number 897481 (HPQC), the Institut für Quanteninformation GmbH, and the EU-QUANTERA project TNiSQ (N-6001).

\end{acknowledgments}

\appendix

\section{Quantum circuit details}\label{app:CircuitDetails}
\begin{figure}
\centering
\subfloat[Three qubits\label{subfig:BulkUnitary_qubits}]
{\includegraphics[width=\columnwidth]{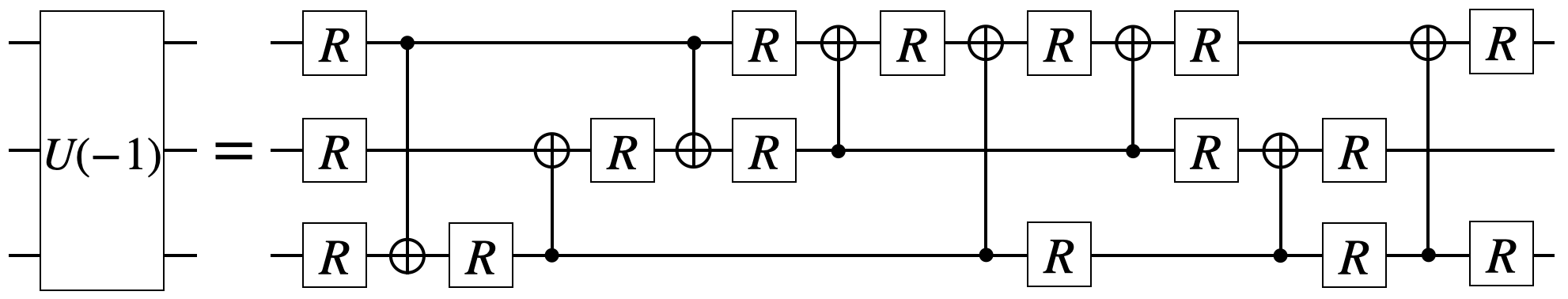}}\\[0.2cm]
\subfloat[Two qutrits\label{subfig:BulkUnitary_qutrits}]
{\includegraphics[width=0.6\columnwidth]{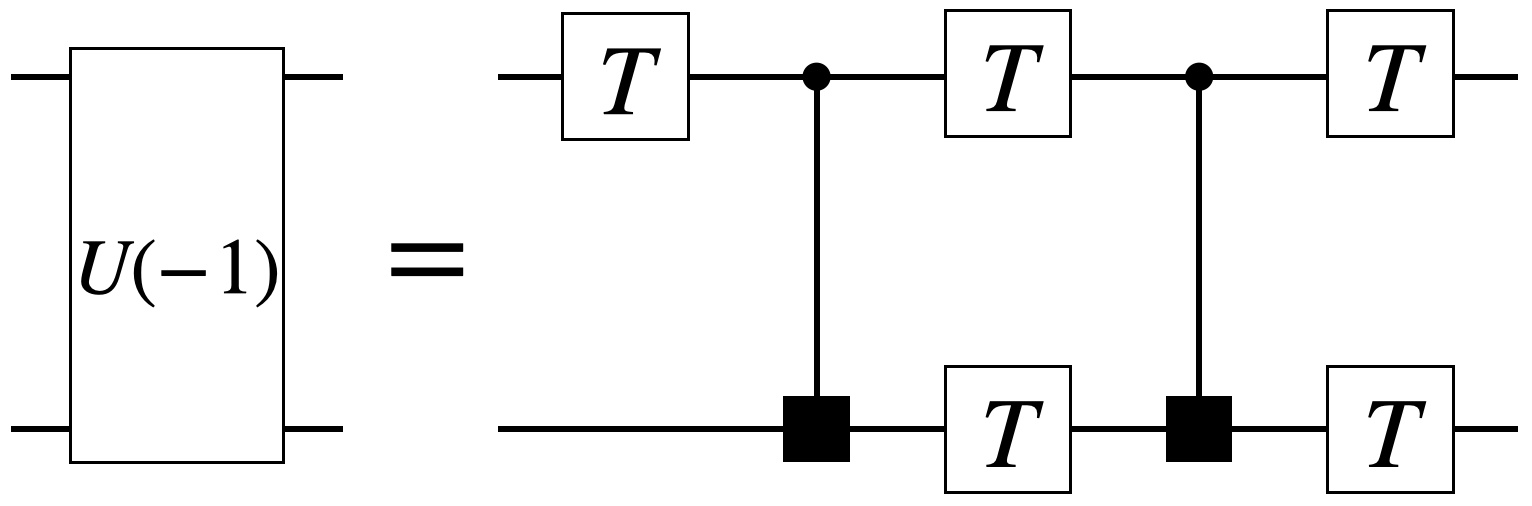}}
\caption{Circuit layout for the conversion of an isometry.
% to a full unitary with near-optimal gate count.
The single qudit operations $R$ and $T$ represent general single qubit and single qutrit rotations, respectively. The entangling two-qutrit CINC gate is represented by a controlled black square.}
\label{fig:BulkUnitaryDecomposition}
\end{figure}

The unitarization process will be illustrated by means of an example. Consider the state at $g=-1$ from the spin-$1$ family in \cref{eq:qutrit-mps}.
After right-normalization, the transformed MPS tensor reads
\begin{equation}
\label{eq:qutrit-mps-right-normalized}
\tilde{B}_+ =
\frac{1}{\sqrt{6}}
\begin{pmatrix}
1 & i \\
i & -1
\end{pmatrix},\
\tilde{B}_\circ =
\frac{1}{\sqrt{3}}
\begin{pmatrix}
0 & 1 \\
-1 & 0
\end{pmatrix},\
\tilde{B}_- =
\tilde{B}_+^\ast.
\end{equation}
The MPS is converted in a unitary operator, using the procedure described in \cref{fig:Unitarization}.
The corresponding three-qubit unitary is then
\begin{equation}
U(g = -1) = \frac{1}{\sqrt{6}}
\begin{pmatrix}
0 & \sqrt{2} & \\
1 & i & \\
1 & -i & \\
0 & 0  & \dots \\
-\sqrt{2} & 0 & \\
i  & - 1 & \\
-i  & - 1 & \\
0 & 0 &
\end{pmatrix}
,
\end{equation}
and the corresponding two-qutrit version is
\begin{equation}
U(g = -1) = 
\frac{1}{\sqrt{6}}
\begin{pmatrix}
0 & 0 & \\
0 & 0 & \\
0 & 0 & \\
0 & \sqrt{2} & \\
1 & i  & \cdots \\
1 & -i  & \\
-\sqrt{2} & 0 &  \\
i  & - 1 & \\
-i  & - 1 & 
\end{pmatrix}
.
\end{equation}
The dots correspond to additional columns, orthogonal to the first ones, which need to be added to obtain a unitary operator~\cite{Iten21,Younis21}.
Decomposing the three-qubit unitary in terms of a standard universal gate set consisting of general three-parameter single-qubit rotations and the CNOT entangling gate by means of the column-by-column decomposition scheme~\cite{Iten16} yields the circuit depicted in Fig.~\ref{subfig:BulkUnitary_qubits}.
For decomposing the three-qutrit unitary, we employ the universal gate set consisting of general single qutrit rotations with eight parameters and the entangling CINC gate, which can be seen as a generalization of the CNOT gate to higher dimensions.
In dimension $d=3$ and using the notation $+,\circ,- \equiv 0,1,2$, its action is defined as
\begin{equation}
\mathrm{CINC}:
\begin{cases*}
\ket{q,k} \mapsto \ket{q,(k + 1)\ \mathrm{mod}\ 3}& \text{if} q = 2 \\
\ket{q,k} \mapsto \ket{q,k} & \text{otherwise}\, ,
\end{cases*}
\end{equation}
incrementing the state of the target qutrit if the control qutrit is in the state $\ket{2}$. For this gate set, the two-qutrit isometry decomposition is obtained by a numerical optimization technique based on instantiation of parametrized circuit templates~\cite{Szasz23}. The layout of the resulting optimized circuit is displayed in Fig.~\ref{subfig:BulkUnitary_qutrits}.\\
Besides the isometries preparing the quantum states, the measurement unitaries $V_i$ also contribute to the total gate count of the circuits. They consist of merely a single qudit rotation in case of the spin-$1/2$ family and the spin-$1$ family implemented on the qutrit platform, but require three entangling CNOT gates and several single qubit rotations in case of the spin-$1$ family realized on the qubit platform. Recall that $V_i$ can also equal the identity, not requiring any gates. For different values of $g$ the gate count can vary. For instance, the environment unitary $U_1$ of the cluster Ising model path requires a single CNOT gate for $g\leq 0$ but two CNOTs otherwise. The bulk isometry $U$ has two CNOT gates for all values of $g$. For a system size of $5$ physical qubits and two ancillary ones, this amounts to a total of $11-12$ CNOT gates for the spin-$1/2$ family of states. Meanwhile, on the qubit platform, the spin-$1$ family requires a single CNOT for the environment unitary, $9$ CNOT gates for the bulk unitary, and $3$ CNOT gates for any measurement unitary that is not the identity, yielding a total of $37$ entangling gates for the circuits of Fig.~\ref{subfig:CircuitAKLTqubits}, if $V_i\neq I \,\forall i$. Only at the exact AKLT point $g=-1$, the bulk unitary requires one fewer CNOT gate ($8$). Finally, on the qutrit platform, the circuits contain a single CINC gate for the environment unitary and $2-3$ CINC gates for the bulk unitary, resulting in a total of $7$ or $10$ entangling gates, depending on the value of $g$.

\section{Circuit transpilation} \label{app:qiskit}
The circuits given in sections~\ref{sec:StatePrep} and~\ref{sec:measurement} were submitted to the qubit setup via our custom modification of the Qiskit backend by Alpine Quantum Technologies (AQT) \cite{aqt_provider}. The customized provider has a transpilation stage to adapt a circuit for running on the experimental setup to enhance performance. The transpilation procedure consists of several stages:
\begin{enumerate}
    \item Using Qiskit default transpiler with optimization level 3 \cite{qiskit_transpiler} to simplify the circuit and convert all the gates to the setup's native gates set: $R(\theta, \phi) = \exp(-i\frac{\theta}{2}(X \cos \phi + Y \sin \phi))$ and $XX(-\pi/2)$.
    \item Minimizing the number of single-qubit gates by combining them and commuting gates through $XX$ gates where possible according to \cite{maslov2017}.
    \item Discarding all $R(\theta, \phi)$ gates with $|\theta| < 10^{-3}$. This is motivated by the fact that the infidelity of a physical single-qubit gate in the setup is on the order of $10^{-4}$, so performing a gate physically will not be better than skipping it.
    \item Replacing $R(\theta, \phi)$ gates with $|\theta| < \pi/15$ with two gates: $R(\theta+\pi, \phi)$ and $R(-\pi, \phi)$. The typical $\pi$-pulse time in the setup is around \SI{30}{\micro\second}, while the minimal allowed pulse time is \SI{1.5}{\micro\second}, which motivates this step.
\end{enumerate}
Running this transpilation procedure yields a circuit with fewer gates than the initial circuit, and thus a higher fidelity measurement result.

\section{Classical shadow details} \label{app:ClassicalShadow}
Here we show in greater detail how the estimators in Eq.~\eqref{eq:featureVector} are computed from the MUB samples. We use a technique called shadow tomography~\cite{Aaronson2018,Aaronson2019}, which is designed specifically for efficient estimation of a set of observables, rather than reconstruction of the full density matrix. We consider a set of $N_s$ MUB samples $\{M_1^{(l)},M_2^{(l)},\dots,M_N^{(l)}\}_{l=1}^{N_s}$ of a $N$-site state $\rho$. Each outcome $M^{(l)}_j$ is a projector onto one of the states from \eqref{eq:MUBstates_d=2} or \eqref{eq:MUBstates_d=3}, acting on site $j$. Sampling MUB states defines a quantum channel on $\rho$ in terms of the expectation value
\begin{equation}
Q_N(\rho) = \lim_{N_s\rightarrow \infty} \frac{1}{N_s} \sum_{l=1}^{N_s} \bigg[\bigotimes_{j=1}^N M^{(l)}_j \bigg] = \mathbb{E}\bigg(\bigotimes_{j=1}^N M_j \bigg),
\end{equation}
which is invertible if the measurements are informationally complete. Inverting this quantum channel yields the approximate expression $\rho \approx \frac{1}{N_s} \sum_{l=1}^{N_s} \hat{\rho}^{(l)}$ for finite $N_s$ in terms of classical shadow elements~\cite{Huang:2020}
\begin{equation}
\hat{\rho}^{(l)} = Q^{-1}_N\bigg(\bigotimes_{j=1}^N M^{(l)}_j\bigg) = \bigotimes_{j=1}^N Q^{-1}_1(M^{(l)}_j).
p
\end{equation}
In the last equality, the inverse quantum channel $Q^{-1}_N$ acting on all $N$ sites must factorize to respect the product structure of the measurements. The inverted single-site quantum channel for a set of $d+1$ MUB in $d$ dimensions is given by
\begin{equation}
Q^{-1}_1(A) = (d+1)A - \mathrm{Tr} (A)I \, ,
\end{equation}
where $A$ denotes an arbitrary Hermitian matrix. Finally, we can compute estimators of observables as
\begin{equation}\label{eq:rank1featureComputation}
\mathrm{Tr}(O_{j_1}^{a_1}\rho) \approx \frac{1}{N_s} \sum_{l=1}^{N_s} \mathrm{Tr} \big(O_{j_1}^{a_1} Q^{-1}_1(M^{(l)}_{j_1})\big),
\end{equation}
at rank 1, and
\begin{multline}\label{eq:rank2featureComputation}
\mathrm{Tr}(O_{j_1}^{a_1} O_{j_2}^{a_2}\rho) \approx \\
\frac{1}{N_s} \sum_{l=1}^{N_s} \mathrm{Tr} \big(O_{j_1}^{a_1} Q^{-1}_1(M^{(l)}_{j_1})\big) \times \mathrm{Tr} \big(O_{j_2}^{a_2} Q^{-1}_1(M^{(l)}_{j_2})\big),
\end{multline}
at rank 2, and so forth.
Outside of the support of the observable all factors are one, since the projectors satisfy $\mathrm{Tr} \big(Q_1^{-1}(M_j^{(l)})\big) = 1$.
For runtime efficiency, the factors $\mathrm{Tr} \big(O_{j}^{a} Q^{-1}_1(M_{j})\big)$ are pre-computed and stored in a $\abs{\mathcal{B}}\times d(d+1)$ lookup table.

As a last remark we note that originally shadow tomography involves a median of means~\cite{Huang:2020} processing of the shadow elements $\{\hat{\rho}^{(l)}\}$. This method of statistical outlier mitigation is omitted in our approach, since TK-SVM already appropriately handles outliers by assigning them a lower (or zero) optimization weight compared to non-outliers.

\section{Comparison of qubit and qutrit implementations}\label{app:QubitQutritComparison}
\begin{figure}
\centering
\includegraphics[width=0.95\columnwidth]{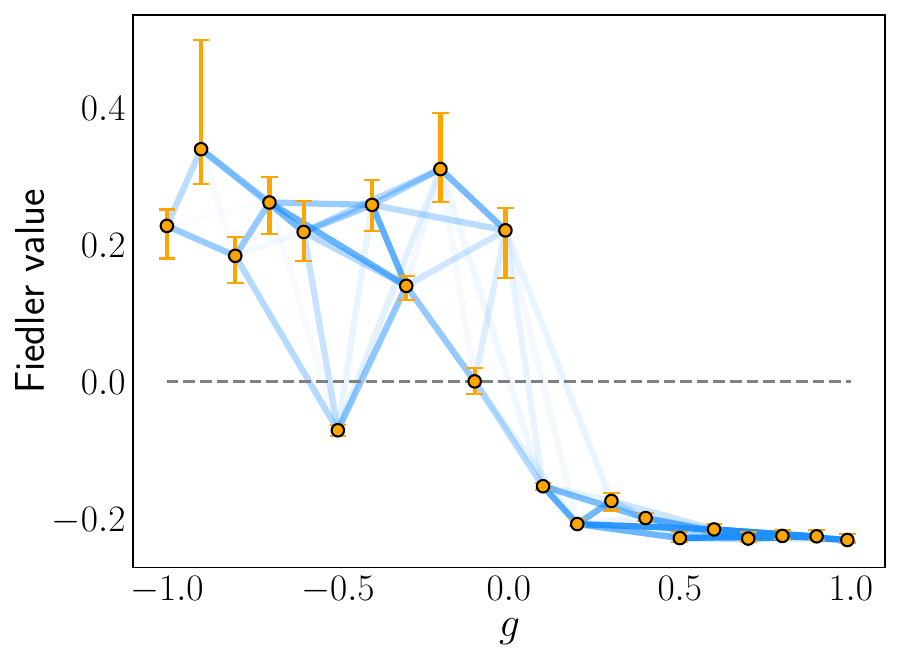}
\caption{Rank $1$ phase classification using qutrits. The sign of the Fiedler value indicates to which part of the graph each vertex belongs. While the part of the graph corresponding to the trivial paramagnetic phase is strongly connected, the part corresponding to the SPT phase is less connected.}
\label{fig:PhasesAKLTqutrits}
\end{figure}
We wish to compare the two different implementations of the AKLT product state path. Starting with the phase classification, and comparing Fig.~\ref{subfig:PhasesAKLTqubits} with Fig.~\ref{fig:PhasesAKLTqutrits}, we see that the qubit implementation yields more consistent results. In case of the qutrit implementation, the trivial phase $g>0$ is represented by one strongly connected part of the graph, as expected, but the part representing the SPT phase $g<0$ is less strongly connected than for the qubits. Still only a single phase point $g=-0.5$ is misclassified and another one $g=-0.1$ has an ambiguous Fiedler value close to zero. Thus the phase classification can still be considered successful and one can proceed with merging the datasets to extract the characteristic features of each phase. At ranks higher than $1$, however, the phase classification algorithm fails to identify two distinct parts of the graph for the qutrit data (therefore not shown), while it still works for the qubit data.\\
Although the qutrit implentation yields slightly less consistent results regarding the phase classification due to lower benchmarking values, it yet performs equally well with regards to identifying the correct local observables characterizing each phase, see Fig.~\ref{fig:aklt-productstate_feat}.

We assume that one of the reasons for the lower consistency of the results produced by the qutrit system is the lower two-qubit gate fidelity as compared to the qubit system.
The qutrit system features two-qubit $XX$ gates with a fidelity of $96.9(1)\%$ \cite{Ringbauer2022}, while the qubit system has a higher two-qubit $XX$ gate fidelity of $98.6(5)\%$ \cite{pogorelov2025thesis}. Moreover, the qubit setup has noticeably lower optical addressing crosstalk values than the qutrit setup \cite{pogorelov2025thesis, erhard2021thesis}.

\begin{figure}
\centering
\subfloat[rank 1]
{\includegraphics[width=0.9\columnwidth]{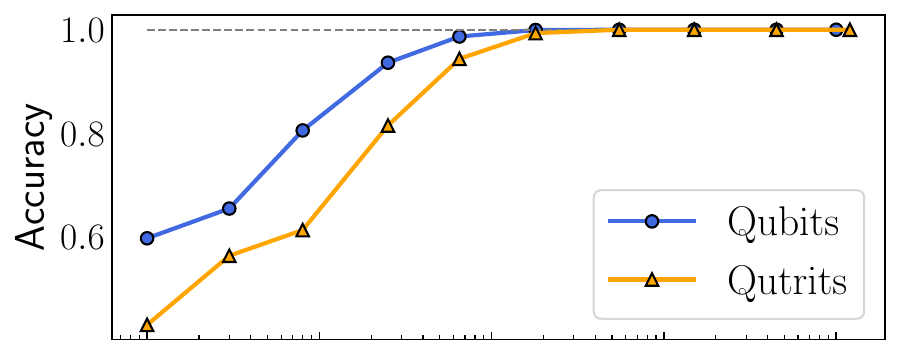}}\\
\subfloat[rank 2]
{\includegraphics[width=0.9\columnwidth]{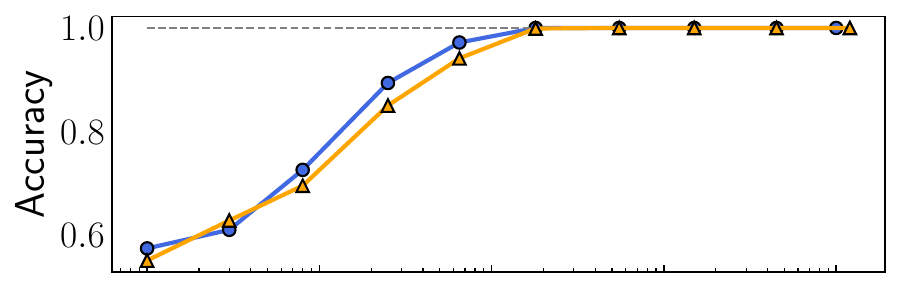}}\\
\subfloat[rank 3]
{\includegraphics[width=0.9\columnwidth]{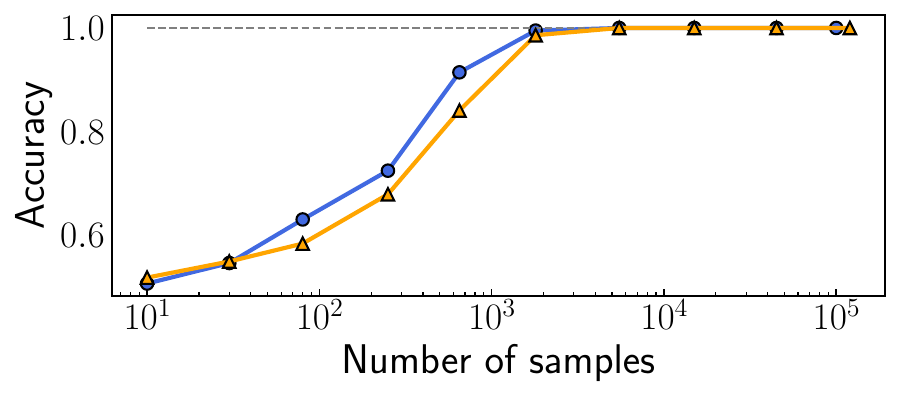}}
\caption{\textbf{Prediction accuracy for the binary classification task between AKLT-state and product-state.} For each classification task, experimental data is used to train the machine, but simulated data is used to test it. The qubit implementation is performing slightly better than the qutrit implementation at rank $1$.}
\label{fig:Accuracy_aklt}
\end{figure}

\section{Accuracy and cluster average}\label{app:Accuracy}
We investigate the prediction accuracy in dependence of the number of samples used for training.
For this purpose we consider the binary classification task for the two datasets that are deepest in each phase of the AKLT product state path, $g=-1$ and $g=+1$.

Once the training is completed, the TK-SVM decision function is used to predict the class of a sufficiently large set of test samples.
The fraction of correctly classified test samples is called accuracy. These test samples are generated in a simulation, therefore free of any noise that might be manifest in the experimental training set.
This approach not only quantifies the minimum number of samples required for successful classification, but also demonstrates the ability of TK-SVM to learn the correct decision boundary from noisy data, see Fig.~\ref{fig:Accuracy_aklt}.
Furthermore, we want to point out that the number of required samples can be drastically reduced if the size of the system is increased.
When computing the feature vector components as exemplified in Eqs.~\eqref{eq:rank1featureComputation} and~\eqref{eq:rank2featureComputation}, the average is taken over several independent samples.
Assuming translational invariance, this average can also be taken over different clusters within the same sample, thereby improving the precision to which the features are estimated.
A system consisting of $L=5$ sites admits merely one single $n=5$-site cluster, whereas for $L=72$ sites there are $L-n+1=68$ overlapping clusters or $\lfloor L/n \rfloor = 14$ non-overlapping clusters.
For more details on the cluster averaging technique, see Appendix B in Ref.~\cite{sadoune23}.

\begin{figure}[h!]
\includegraphics[width=0.9\columnwidth]{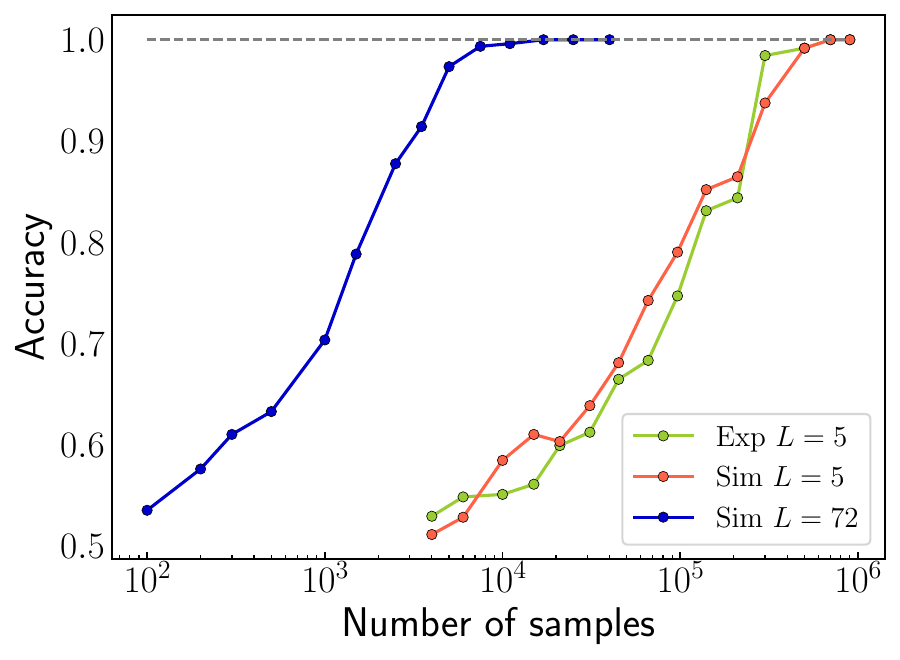}
\caption{\textbf{Prediction accuracy for the cluster Ising model.} The machine is trained at rank $5$ with cluster size $n=5$ in all cases. For physical system size $L=5$ the machine is trained using experiment as well as simulated data, while for system size $L=72$ only simulated data is available for training. The test sets used to determine prediction accuracy is simulated data in all cases.}
\label{fig:Accuracy_clusterModel}
\end{figure}

Again we consider a binary classification task, this time classifying the whole SPT phase against the whole trivial phase by merging datasets with $g<0$ and $g>0$, respectively.
The test set always consists of simulated data, whereas for training we use both simulated and experiment data for different runs.
Our results are displayed in Fig.~\ref{fig:Accuracy_clusterModel}.
As expected, the (overlapping) cluster average reduces the number of required samples by roughly a factor of 68.
Moreover, the machine classifies simulated test samples with comparable accuracy when trained with simulated and experimental data.

\bibliographystyle{unsrtnat}

\bibliography{references}

\end{document}